\documentclass[numberedappendix,twocolappendix,iop]{emulateapj}
\usepackage{amsmath}
\usepackage{amssymb}
\usepackage{bm}
\usepackage{courier}

\begin{document}


\bibliographystyle{apj}

\title{The Arches Cluster: Extended Structure and Tidal Radius}
\author{Matthew W. Hosek Jr.\altaffilmark{1}, Jessica R. Lu\altaffilmark{1}, Jay Anderson\altaffilmark{2}, Andrea M. Ghez\altaffilmark{3}, Mark R. Morris\altaffilmark{3}, William I. Clarkson\altaffilmark{4}}
\altaffiltext{1}{Institute for Astronomy, University of Hawaii, 2680 Woodlawn Drive, Honolulu, HI 96822, USA; mwhosek@ifa.hawaii.edu, jlu@ifa.hawaii.edu}
\altaffiltext{2}{Space Telescope Science Institute, 3700 San Martin Drive, Baltimore, MD 21218, USA; jayander@stsci.edu}
\altaffiltext{3}{UCLA Department of Physics and Astronomy, Los Angeles, CA 90095, USA; ghez@astro.ucla.edu, morris@astro.ucla.edu}
\altaffiltext{4}{Department of Natural Sciences, University of Michigan-Dearborn, 4901 Evergreen Road, Dearborn, MI 48128; wiclarks@umich.edu}

\begin{abstract}

At a projected distance of $\sim$26 pc from Sgr A*, the Arches cluster provides insight to star formation in the extreme Galactic Center (GC) environment. Despite its importance, many key properties such as the cluster's internal structure and orbital history are not well known. We present an astrometric and photometric study of the outer region of the Arches cluster (R $>$ 6.25") using \emph{HST} WFC3IR. Using proper motions we calculate membership probabilities for stars down to F153M = 20 mag ($\sim$2.5 M$_{\odot}$) over a 120" x 120" field of view, an area 144 times larger than previous astrometric studies of the cluster. We construct the radial profile of the Arches to a radius of 75" ($\sim$3 pc at 8 kpc), which can be well described by a single power law. From this profile we place a 3$\sigma$ lower limit of 2.8 pc on the observed tidal radius, which is larger than the predicted tidal radius (1 -- 2.5 pc). Evidence of mass segregation is observed throughout the cluster and no tidal tail structures are apparent along the orbital path. The absence of breaks in the profile suggests that the Arches has not likely experienced its closest approach to the GC between $\sim$0.2 -- 1 Myr ago. If accurate, this constraint indicates that the cluster is on a prograde orbit and is located front of the sky plane that intersects Sgr A*. However, further simulations of clusters in the GC potential are required to interpret the observed profile with more confidence.

\end{abstract}
\maketitle

\section{Introduction}
\label{sec:intro}
The Arches cluster is a young (2-4 Myr; \citeauthor{Najarro:2004ij} 2004, \citeauthor{Martins:2008hl} 2008) massive ($\sim$4--6 x 10$^4$ M$_{\odot}$; \citeauthor{Clarkson:2012ty} 2012) star cluster near the center of the Milky Way. It has a projected distance of just $\sim$26 pc from the supermassive black hole (SMBH) and is one of the most centrally concentrated star clusters in the Galaxy. Old enough to be free of its natal gas cloud and yet young enough to sample the full stellar mass range, the Arches cluster provides a unique opportunity to probe star formation and cluster evolution in the extreme Galactic Center (GC) environment. However, studies of the cluster are complicated by stellar crowding and the high level of extinction which varies significantly across the field \citep[1.6 mag $<$ A$_{Ks} <$ 3.3 mag;][]{Habibi:2013th}. This effectively smears out the photometric properties of the cluster population, making it difficult to separate cluster members from field stars through photometry alone. As a result, important questions about the cluster's structure, initial mass function, and orbital history remain.

The Arches is one of the closest examples of a young massive cluster (YMC) in a strong tidal field. Such objects are not predicted to have long lifetimes, as simulated clusters near the GC show complete tidal disruption on the order of $\sim$10 Myr \citep{Kim:1999fk, Kim:2000wd} or shorter when interactions with giant molecular clouds are considered \citep{Kruijssen:2014rf}. Since the effects of tidal perturbations are most significant for stars on the outskirts of their clusters \citep{Gnedin:1999lp, Kupper:2010ek}, measuring the structure of the outer region of the Arches offers insight to its past interactions. For example, it was long thought that the observed tidal radius (i.e., limiting radius) of a cluster, where the stellar density drops to zero, should correspond to its theoretical tidal radius (i.e., Jacobi radius), where the gravitational acceleration of the cluster equals the tidal acceleration of its parent galaxy \citep{von-Hoerner:1957qv}. Early Fokker-Planck simulations of clusters on eccentric orbits further suggested that the tidal radius imposed by the strongest tides at perigalacticon should persist to later times \citep{Oh:1992uo}. However, more recent N-body simulations show that perigalacticon passage does not cleanly truncate a cluster, but rather results in an extended radial profile that approaches a power law \citep{Oh:1995xy, Johnston:1999qf, Penarrubia:2009pd, Kupper:2010ek}. The ``extratidal'' stars can have a different profile slope than the rest of the cluster, creating a break in the profile at a radius which may be related to the time since perigalacticon passage \citep{Penarrubia:2009pd, okas:2013hc}.

Measurements of globular cluster profiles have revealed the presence of extratidal stars, often associated with tidal tail structures \citep{Odenkirchen:2001fr, Siegel:2001qq, Odenkirchen:2003jk, Belokurov:2006qf, Sollima:2011yu, Chun:2015fj}. However, as discussed by \citet{Kupper:2010ek} and \citet{Carballo-Bello:2012kx}, even clusters that are not exposed to varying external tidal fields can also form extratidal structures through two-body relaxation, as stars which become energetically unbound from the cluster do not escape instantaneously but rather on a time-scale which is dependent on their orbital parameters \citep{Fukushige:2000lq, Baumgardt:2003qy, Zotos:2015la}. Both of these mechanisms affect the radial profile of globular clusters, which have ages larger than their relaxation times. On the other hand, two-body relaxation is negligible for YMCs that are much younger than than their relaxation times, and so the impact of tidal perturbations on their profiles should be easier to isolate. Given the high likelihood of strong tidal interactions with the GC, the Arches offers a promising opportunity to measure such a feature. In addition, a detailed understanding of the Arches cluster profile is necessary for assessing the impact of dynamical effects on the present-day mass function (i.e. mass segregation and tidal stripping).

Insight into the past tidal interactions of the Arches cluster also provides valuable information about its orbit. While its bulk proper motion \citep{Stolte:2008qy, Clarkson:2012ty} and doppler velocity \citep{Figer:2002nr} have been measured, the line-of-sight distance is unknown, preventing a unique orbital solution. As a result, the birth environment of the cluster and its relation to the nearby Quintuplet cluster are not well understood. For example, it has been suggested that both the Arches and Quintuplet formed in a collision between gas clouds along the X$_1$ and X$_2$ orbit families in the Galactic bar, which may be a region of highly efficient star formation \citep{Binney:1991cr, Stolte:2014qf}. Alternatively, the Arches may be the end product of a cluster formation sequence identified by \citet{Longmore:2013lq} and \citet{Kruijssen:2015fx}, where starburst clusters form from the tidal compression of gas clouds that pass close to the GC. It has also been suggested that the Arches may be a possible source of the isolated massive stars observed near the GC, depending on its orbit and how much tidal stripping has occurred \citep{Mauerhan:2010qv, Habibi:2014qf}. If the time since the last tidal perturbation (presumably occurring at the closest approach to the GC) can be established, the set of possible orbits for the Arches calculated by \citet{Stolte:2008qy} can be significantly restricted.

We present the radial profile of the Arches cluster out to large cluster radii ($\sim$75'', or $\sim$3 pc at 8 kpc\footnote{All distances throughout this paper assume a distance of 8 kpc to the Arches cluster.}) and investigate the structure of the outer region of the cluster for the first time. We use stellar proper motions rather than photometry to calculate cluster membership probabilities, avoiding many of the difficulties introduced by differential reddening. The effectiveness of this method on the Arches cluster was demonstrated by \citet{Stolte:2008qy} and \citet{Clarkson:2012ty}, who used ground-based adaptive optics (AO) observations to measure the cluster's bulk proper motion and identify members in the central 10" x 10" region. In this paper, we conduct an astrometric study of the Arches cluster using the \emph{Hubble Space Telescope} (HST), which provides high astrometric precision over a field of view 144 times larger than these previous studies. This allows us to measure the radial profile to beyond the predicted tidal radius \citep[25" - 60";][]{Kim:2000wd, Portegies-Zwart:2002hc}. In addition, we examine the degree of mass segregation throughout the cluster and search for the presence of tidal tails. The consequences of our results are discussed in relation to the orbital history of the Arches cluster.

\section{Methods}
\subsection{Observations and Measurements}
\label{sec:obs}
\begin{figure}
\begin{center}
\includegraphics[scale=0.40]{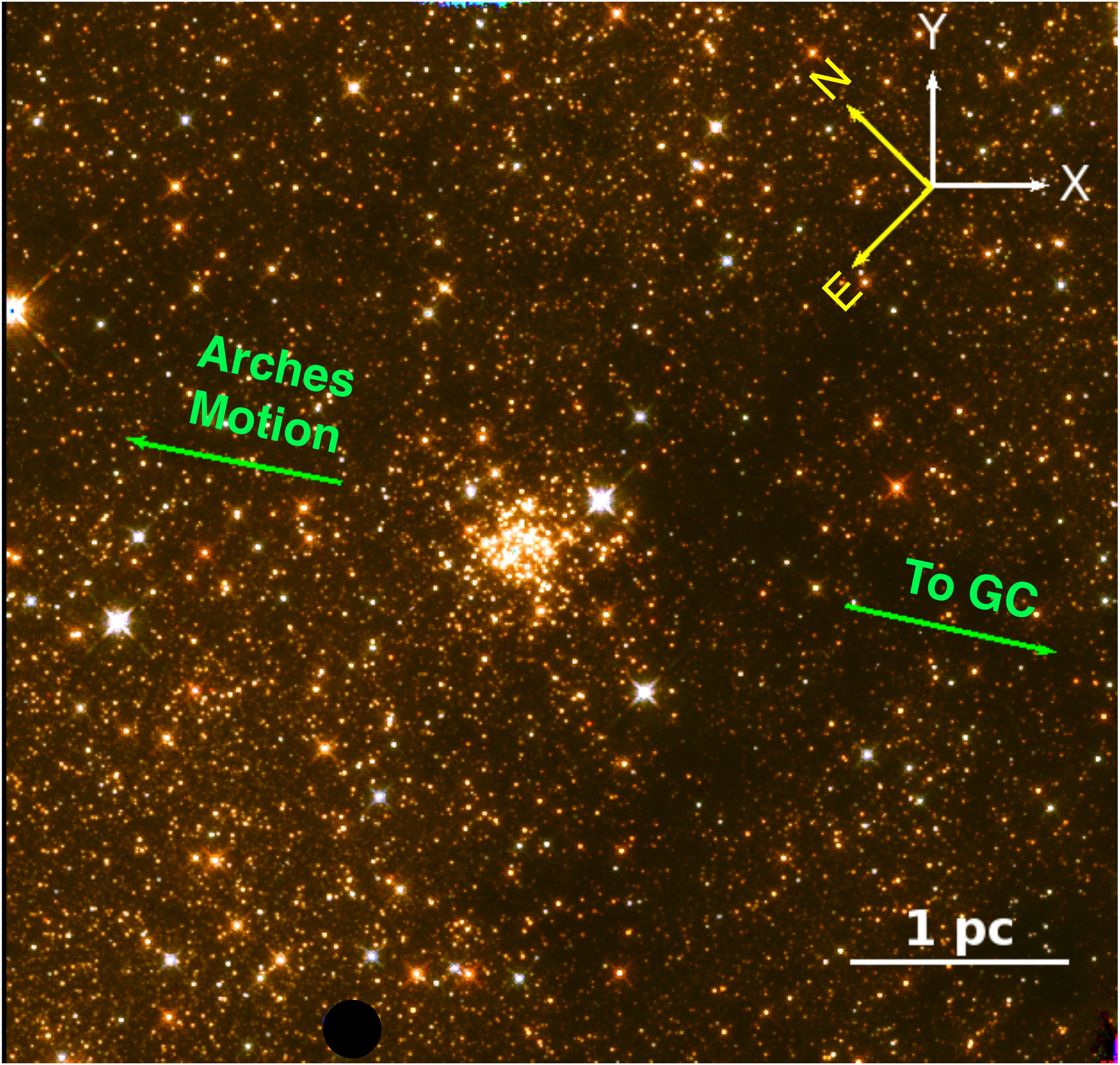}
\caption{Three color image of the Arches Cluster, with F127M~=~blue, F139M = green, and F153M = red. Significant differential extinction is apparent from the changing density of the field stars. The hole in the lower left side of the image is due to a known defect in the WFC3IR chip. The proper motion of the cluster (labeled with green arrow) is very nearly parallel to the Galactic plane.}
\label{Arches_color}
\end{center}
\end{figure}

We observed the Arches cluster with the \emph{Hubble Space Telescope} (HST) WFC3IR camera using the F127M, F139M, and F153M filters (1.27 $\mu$m, 1.39 $\mu$m, and 1.53 $\mu$m, respectively; PI: Ghez, ID: 11671, 12318, 12667). A summary of the observations is provided in Table \ref{ObsTable}. These observations have a field of view of 120'' x 120'' and are centered at $\alpha$(J2000) = 266.4604, $\delta$(J2000)~=~-28.8222 with a position angle of -45$^{\circ}$ (Figure \ref{Arches_color}). Astrometry is performed on the F153M observations, which were obtained in three epochs over a two year baseline between 2010 -- 2012. Of this filter set, F153M was chosen for astrometry because it provides the optimal combination of limited saturation and a well-sampled point spread function (PSF) with a FWHM $\sim$0.17" (1.4 pix, scale = 0.121" pix$^{-1}$).  High astrometric and photometric precision is achieved by observing at the same position angle and pixel position across epochs and using a dense, sub-pixel spiral dithering pattern within each epoch. F127M and F139M observations were only obtained in 2010 using a simpler dither pattern for the purpose of color information to derive extinction.

\begin{deluxetable*}{c c c c c c c}
\tablewidth{0pt}
\tabletypesize{\footnotesize}
\tablecaption{HST WFC3IR Observations}
\tablehead{
\colhead{Date} & \colhead{Filter} & \colhead{N$_{images}$} & \colhead{Total Exp. Time (s)} & \colhead{Depth (mag)\tablenotemark{b}} & \colhead{Pos Error (mas)\tablenotemark{a}} & \colhead{Phot Error (mag)\tablenotemark{a}}
}
\startdata
2010.6150 & F127M & 12 & 7200 & 23.63 & 0.60 & 0.02  \\
2010.6148 & F139M & 10 & 3500 & 23.29 & 0.90 &  0.03 \\
2010.6043 & F153M & 21 & 7350 & 23.31 & 0.88  & 0.05 \\
2011.6829 & F153M & 21 & 7350  & 23.32 & 0.88 & 0.05 \\
2012.6156 & F153M & 21 & 7350 & 23.31 & 0.88 & 0.05
\enddata
\tablenotetext{a}{Median value at mag = 20 in respective filter}
\tablenotetext{b}{Estimated from 95th percentile of extracted stellar magnitudes.}
\label{ObsTable}
\end{deluxetable*}

Each frame is processed using the standard \emph{HST} pipeline, which produces a \texttt{flt} image which has been flat-fielded and bias-subtracted. We use a combination of public and custom software to extract high-precision astrometry and photometry via PSF-fitting with a variable PSF model. While this method has been previously implemented on HST optical observations of Globular Clusters, this is the first time it has been applied to WFC3IR observations. We produce a list of individual stellar measurements for each filter/epoch using the program \emph{KS2}, a generalization of the software developed to reduce the Globular-Cluster Treasury Program \citep{Anderson:2008qy}. Star positions are transformed to an arbitrary astrometric reference frame where the net motion of the cluster plus field is 0 mas yr$^{-1}$ using general 6-parameter linear transformations that can be described as a 2D translation, rotation, plate scale, and shear for each image. Approximately $\sim$50,000 stars are measured in each filter/epoch. The photometry is calibrated to the standard Vega magnitude system using the zero points derived for the WFC3IR camera\footnote{As of August 2014; http://www.stsci.edu/hst/wfc3/phot\_zp\_lbn}. A detailed description of the data reduction and measurement process, as well as an analysis of the astrometric and photometric errors, is provided in Appendix \ref{sec:Obs_methods_appendix}.

We calculate proper motions for the stars that are detected in all three F153M epochs, using linear fits weighted by the individual astrometric errors. Proper motion uncertainty as a function of observed F153M magnitude is presented in Figure \ref{velerr}. Several tests were conducted to confirm the validity of these errors (see Appendix \ref{sec:Obs_methods_appendix}). Previous studies of the bulk proper motion of the Arches cluster relative to the field population by \citet{Stolte:2008qy} and \citet{Clarkson:2012ty} revealed that a precision of $\sim$0.8 mas yr$^{-1}$ is needed in order to reliably separate cluster members from field stars. As a conservative error cut, we restrict the forthcoming analysis to stars with a proper motion precision of 0.65 mas yr$^{-1}$ or better, which we achieve down to F153M $\approx$ 20 mag. At the average distance and reddening of the Arches cluster this corresponds to roughly 2.5 M$_{\odot}$. We additionally require a minimum photometric precision of 0.06 mags in each F153M epoch to ensure high-quality results. We measure $\sim$26,000 proper motions, $\sim$6000 of which pass these error cuts. The kinematic distinction between cluster and field stars is clearly seen in a vector point diagram (VPD; Figure \ref{VPD}). Note that the proper motions have been rotated from image coordinates into projected equatorial coordinates where the two-dimensional proper motion vector of any star is $\bm{\mu}$~=~[$\mu_{\alpha}$cos$\delta$, $\mu_{\delta}$]. Proper motions are also shifted into a reference frame where the cluster is at rest, estimated from the mean motion of stars within the central 10"x10" region of the cluster. This sample is an order of magnitude larger than the sample analyzed by \citet{Clarkson:2012ty}, who had a much smaller field of view.

\begin{figure}
\begin{center}
\includegraphics[scale=0.35]{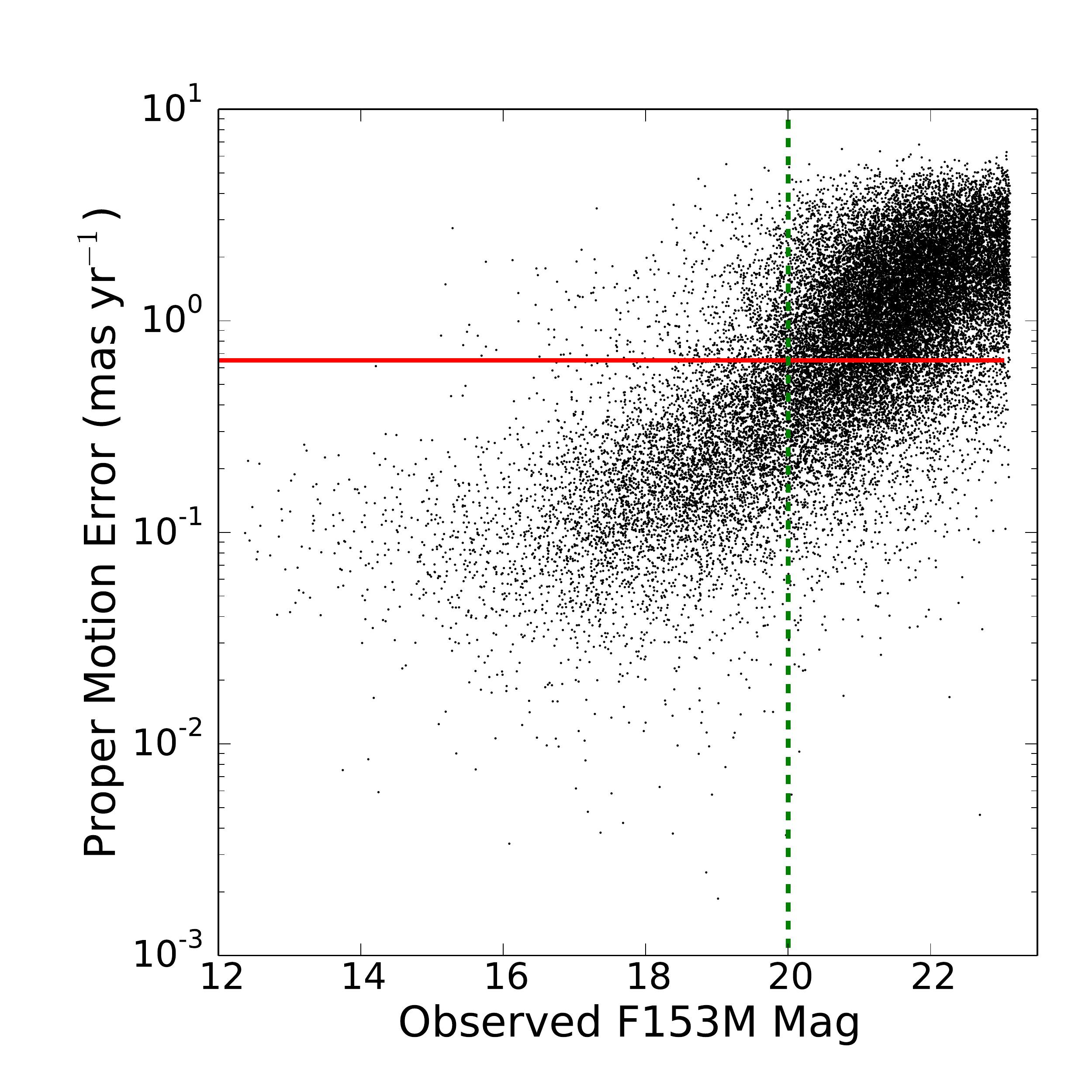}
\caption{Proper Motion error vs. F153M magnitude. The red line marks our proper motion error cut of 0.65 mas yr$^{-1}$. Only stars below this cut are included in our analysis. The green line denotes F153M = 20 mag, which corresponds to $\sim$2.5 M$_{\odot}$ at the approximate distance and average reddening of the Arches cluster. }
\label{velerr}
\end{center}
\end{figure}

\begin{figure}
\begin{center}
\includegraphics[scale=0.35]{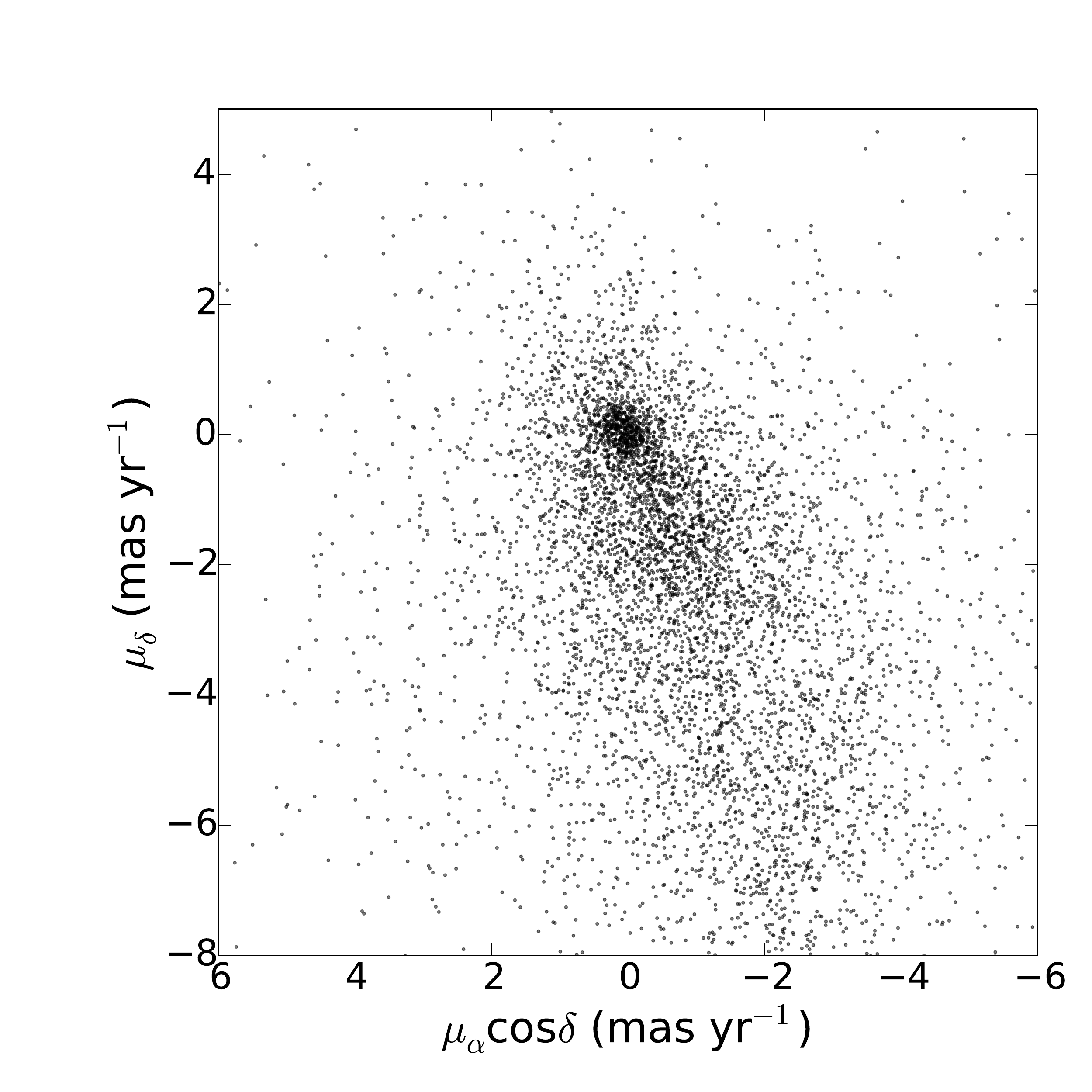}
\caption{Vector point diagram of the $\sim$6000 stars included in our analysis. Proper motions are in the reference frame of the cluster. Cluster members appear as a distinct clump of stars at ($\mu_{\alpha}$cos$\delta$, $\mu_{\delta}$)~=~(0,~0), while the field stars are spread along the Galactic plane.}
\label{VPD}
\end{center}
\end{figure}

\subsection{Cluster and Field Populations}
\label{sec:pops}
In order to calculate cluster membership probabilities we must first characterize the kinematic distributions of the cluster and field star populations. Previous studies of the Arches cluster have assumed that the field kinematics can be modeled as a single elliptical Gaussian distribution \citep{Clarkson:2012ty}. The field kinematic distribution is elliptical because it is primarily composed of stars in the Galactic bulge that exhibit a larger velocity dispersion along the Galactic plane than perpendicular to it, a consequence of coherent rotation \citep{Clarkson:2008hw, Howard:2009ad, Kunder:2012pt}. However, close inspection of Figure \ref{VPD} reveals that the field population cannot be described by a single Gaussian function. To account for this complexity, we adopt a Normal Mixture Model \citep{McLaughlinPeel2000} to simultaneously fit multiple Gaussians to the observed VPD. A more complete description of the field requires modeling the stellar density, kinematics, and reddening at all distances along our line of sight towards the Arches. Such analysis would be valuable for exploring Galactic structure but is beyond the scope of this paper.

We construct a likelihood function for each star in the sample from from the sum of $K$ Gaussian components:

\begin{equation}
\label{likelihood}
L(\bm{\mu}_i) = \sum^{K}_{k=0} \pi_{k} \frac{1}{2\pi |\bm{\Sigma}_{ki}|^{1/2}} exp\left( -\frac{1}{2}(\bm{\mu}_i - \bm{\overline{\mu}}_k)^T \bm{\Sigma}_{ki}^{-1} (\bm{\mu}_i - \bm{\overline{\mu}}_k)\right)
\end{equation}
\begin{equation*}
\mathcal{L} =  \prod_i^N L(\bm{\mu}_i)
\end{equation*}
\\
where $\bm{\mu}_i$ is the proper motion of the $i$th star, $\pi_{k}$ is the fraction of total stars in the $k$th Gaussian such that $\sum_{k=0}^K \pi_k = 1$, $\bm{\overline{\mu}}_k$ is the velocity centroid of the $k$th Gaussian, and $\bm{\Sigma}_{ki}$ is the covariance of the $k$th Gaussian and $i$th star. The total likelihood over the sample of $N$ stars is $\mathcal{L}$. Following \citet{Clarkson:2012ty}, we add the covariance matrices of the population model and stellar proper motion uncertainties such that $\bm{\Sigma}_{ki} = \bm{S}_i + \bm{Z}_k$, where $\bm{S}_i$ is the velocity error matrix (assumed to be diagonal with velocity error components $\sigma_{\mu_{\alpha cos\delta}}^2$ and $\sigma_{\mu_\delta}^2$) and $\bm{Z}_k$ is the covariance matrix of the $k$th Gaussian fit.

With the likelihood function defined, we can determine the global kinematic parameters of the cluster and field populations through Bayesian inference using Bayes' theorem:

\begin{equation}
P(\bm{\pi}, \bm{\overline{\mu}},  \bm{Z} | \bm{\mu}, \bm{S}) = \frac{P(\bm{\mu}, \bm{S} | \bm{\pi}, \bm{\overline{\mu}}, \bm{Z}) P(\bm{\pi}, \bm{\overline{\mu}}, \bm{Z})}{P(\bm{\mu}, \bm{S})}
\end{equation}
\\
where $P(\bm{\pi}, \bm{\overline{\mu}}, \bm{Z} | \bm{\mu}, \bm{S})$ is the posterior probability of our model parameters $\bm{\pi}$, the set of $\pi_k$ values; $\bm{\overline{\mu}}$, the set of Gaussian velocity centroids; and $\bm{Z}$, the set of Gaussian covariance matrices given the observed stellar velocities $\bm{\mu}$ and velocity error matrix $\bm{S}$. $P(\bm{\mu}, \bm{S} | \bm{\pi}, \bm{\overline{\mu}}, \bm{Z})$ is the probability of the observed stellar velocity distribution given the model, and $P(\bm{\pi}, \bm{\overline{\mu}}, \bm{Z})$ is the prior probability of the model. In this case, $P(\bm{\mu}, \bm{S} | \bm{\pi}, \bm{\overline{\mu}}, \bm{Z})$ is the total likelihood $\mathcal{L}$ defined in Equation \ref{likelihood}.

To find the posterior probability distribution we use \emph{Multinest}, a publicly available nested sampling algorithm which serves as an alternative to Markov Chain Monte Carlo (MCMC) algorithms when exploring multi-modal parameter spaces \citep{Feroz:2009lq}. This iterative technique calculates the posterior probability at a fixed number of points in the parameter space and identifies possible peaks, restricting subsequent sampling to the regions around these peaks until the change in evidence drops below a user-defined tolerance level. Multiple peaks can be identified and evaluated, resulting in increased sampling efficiency with complicated parameter spaces. We run the algorithm using the python module \emph{PyMultinest} \citep{Buchner:2014wa}.

\begin{figure*}
\begin{center}
\includegraphics[scale=0.4]{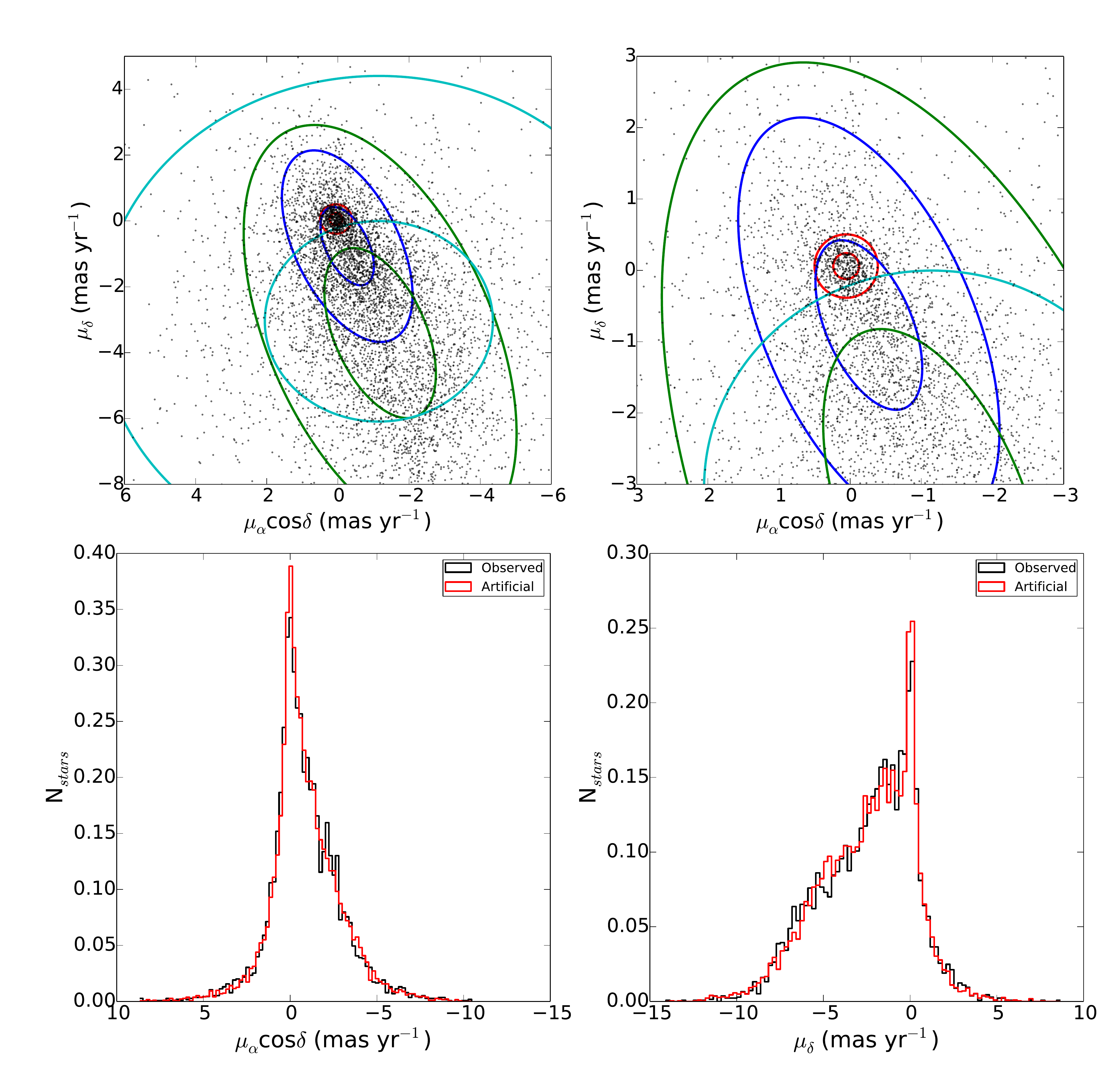}
\caption{\emph{Top:} Vector point diagram of our sample with the fitted 1- and 2$\sigma$ distributions of the cluster Gaussian (red) and field Gaussians (blue, green, cyan corresponding to field Gaussians 1, 2, and 3 in Table \ref{PopModel}, respectively). The \emph{left} plot shows all stars in the field, demonstrating the extension of the field populations in the direction of the Galactic plane. The \emph{right} plot is a zoomed-in view of the cluster population, readily apparent as a tight clump of stars moving with a common motion relative to the field. \emph{Bottom:} Proper motion distribution of the stars in projected equatorial coordinates (\emph{left}: RA, \emph{right}: Dec). The predicted distribution of the Normal Mixture Model (red) is found to be a good match to the observed stars (black).}
\label{VPD_fit}
\end{center}
\end{figure*}

We find that the cluster and field populations can be well described with a 4-Gaussian mixture model, with one Gaussian describing the cluster and the other three describing the field (Figure \ref{VPD_fit}). The use of this model is justified by the Bayesian Information Criterion \citep{Schwarz+78}, which is minimized compared to less complicated (3-Gaussian) or more complicated (5-Gaussian) mixture models. We require the cluster Gaussian to be circular, consistent with the results of \citet{Clarkson:2012ty}, and adopt a prior to roughly constrain its location around (v$_x$, v$_y$) = (0,0). The parameters for the field Gaussians as well as the remaining parameters for the cluster Gaussian are unconstrained. The one-dimensional posterior distributions are well described by a Gaussian function, which is used to determine the best-fit value and error for that parameter. A summary of the parameter priors, best-fit results, and errors is provided in Table \ref{PopModel}.

These results can be compared with those of \citet{Clarkson:2012ty}, who examine the kinematics of the inner 10" x 10" of the cluster using ground-based AO observations. Their measurements have a higher precision but much smaller field of view than our observations. We obtain a velocity dispersion of 0.18 $\pm$ 0.02 mas yr$^{-1}$ for the cluster, which is consistent with the measurement of 0.15 $\pm$ 0.01 mas yr$^{-1}$ by \citet{Clarkson:2012ty} within errors. This agreement comes despite using fully independent data sets which focus on different regions of the cluster.

However, there is less agreement on the bulk motion of the Arches relative to the field population. \citet{Clarkson:2012ty} model the field using a single elliptical Gaussian distribution with a velocity center offset by 4.39 $\pm$ 0.38 mas yr$^{-1}$ from the cluster. In this study we model the field using 3 elliptical Gaussians, and our fits indicate that all have smaller motions relative to the cluster than the \citet{Clarkson:2012ty} result (see Table \ref{PopModel}). If we calculate the average motion of the field from the average sum of the 3 field Gaussians, then we get an overall field motion of 2.83 $\pm$ 0.33 mas yr$^{-1}$ relative to the cluster, which is substantially and significantly lower than the \citet{Clarkson:2012ty} result. We discuss this discrepancy further in $\mathsection$\ref{sec:Profile_implications}.

\begin{deluxetable*}{c |c c |c c | c c | c c}
\tabletypesize{\footnotesize}
\tablecaption{Cluster and Field Population Model: Free Parameters, Priors, and Results}
\tablehead{
& \multicolumn{2}{c}{Cluster Gaussian} & \multicolumn{2}{c}{Field Gaussian 1} & \multicolumn{2}{c}{Field Gaussian 2} & \multicolumn{2}{c}{Field Gaussian 3} \\
\colhead{Parameter\tablenotemark{a}} & \colhead{Prior\tablenotemark{b}} & \colhead{Result} & \colhead{Prior} & \colhead{Result} & \colhead{Prior} & \colhead{Result} & \colhead{Prior} & \colhead{Result}
}
\startdata
$\pi_{k}$ & U(0, 1) & 0.08 $\pm$ 0.01 & U(0, 1) & 0.25 $\pm$ 0.03 & U(0, 1) & 0.42 $\pm$ 0.04 & U(0, 1) & 0.25 $\pm$ 0.03  \\
$\mu_{\alpha, k}$ (mas yr$^{-1}$) & G(0, 0.2) & 0.06 $\pm$ 0.03 & U(-4, 12) & -0.26 $\pm$ 0.06 & U(-4, 12) & -1.19 $\pm$ 0.11 & U(-4, 12) & -1.15 $\pm$ 0.15 \\
$\mu_{\delta, k}$ (mas yr$^{-1}$) & G(0, 0.2) & 0.06 $\pm$ 0.02 & U(-4, 12) & -0.77 $\pm$ 0.09 & U(-4, 12) & -3.40 $\pm$ 0.25 & U(-4, 12) & -3.05 $\pm$ 0.20 \\
$\sigma_{a, k}$ (mas yr$^{-1}$) & U(0, 8) & 0.18 $\pm$ 0.02 & U(0, 8) & 1.27 $\pm$ 0.09 & U(0, 8) & 2.73 $\pm$ 0.12 & U(0, 8) & 3.20 $\pm$ 0.13 \\
$\sigma_{b, k}$ (mas yr$^{-1}$) & $\sigma_b$ = $\sigma_a$ & 0.18 $\pm$ 0.02 & U(0, 4) & 0.60 $\pm$ 0.05 & U(0, 4) & 1.30 $\pm$ 0.08 & U(0, 4) & 3.05 $\pm$ 0.13 \\
$\theta_{k}$ (rad) & --- & 0 & U(0, $\pi$) & 1.16 $\pm$ 0.05 & U(0, $\pi$) & 1.20 $\pm$ 0.03 & U(0, $\pi$) &  0
\enddata
\label{PopModel}
\tablenotetext{a}{Description of parameters: $\pi_k$ = fraction of stars in Gaussian; $\mu_{\alpha, k}$ = RA-velocity centroid of Gaussian; $\mu_{\delta,k}$ = DEC-velocity centroid of Gaussian; $\sigma_{a,k}$ = semi-major axis of Gaussian; $\sigma_{b,k}$ = semi-minor axis of Gaussian; $\theta_{k}$ = angle between $\sigma_{a,k}$ and the RA-axis}
\tablenotetext{b}{Uniform distributions: U(min, max), where min and max are bounds of the distribution; Gaussian distributions: G($\mu$, $\sigma$), where $\mu$ is the mean and $\sigma$ is the standard deviation}
\end{deluxetable*}

\subsection{Extinction Map Using Red Clump Stars}
\label{sec:extmap}
Taking advantage of the high photometric precision of HST, we use red clump (RC) stars in the Galactic bulge to measure the extinction across the Arches cluster field. This provides an alternative to ``sliding" apparent cluster members along their reddening vector in a color-magnitude diagram (CMD) to a theoretical cluster isochrone to measure extinction, as has been done in previous studies of the Arches cluster \citep{Kim:2006fy, Espinoza:2009bs, Habibi:2013th}. This CMD sliding method is prone to field contamination and isochrone uncertainties, especially for the pre-main sequence at low to intermediate masses. On the other hand, stellar evolution theory and observations show that RC stars exhibit well-defined luminosities and colors which do not vary significantly with age or metallicity \citep{C92, P98, S00}, making them useful calibrators to measure extinction. This is especially true near the GC, where the relatively high density of RC stars in the bulge population makes it possible to create reddening maps of different regions \citep{Sumi:2004rr, Sch10}. Though the line-of-sight position of the Arches with respect to the bulge RC stars is uncertain, we assume that all of the extinction is caused by foreground material and so the RC population exhibits similar extinction as the cluster members themselves. With this approach we create the first RC-based extinction map of the Arches cluster.

\begin{figure*}
\begin{center}
\includegraphics[scale=0.35]{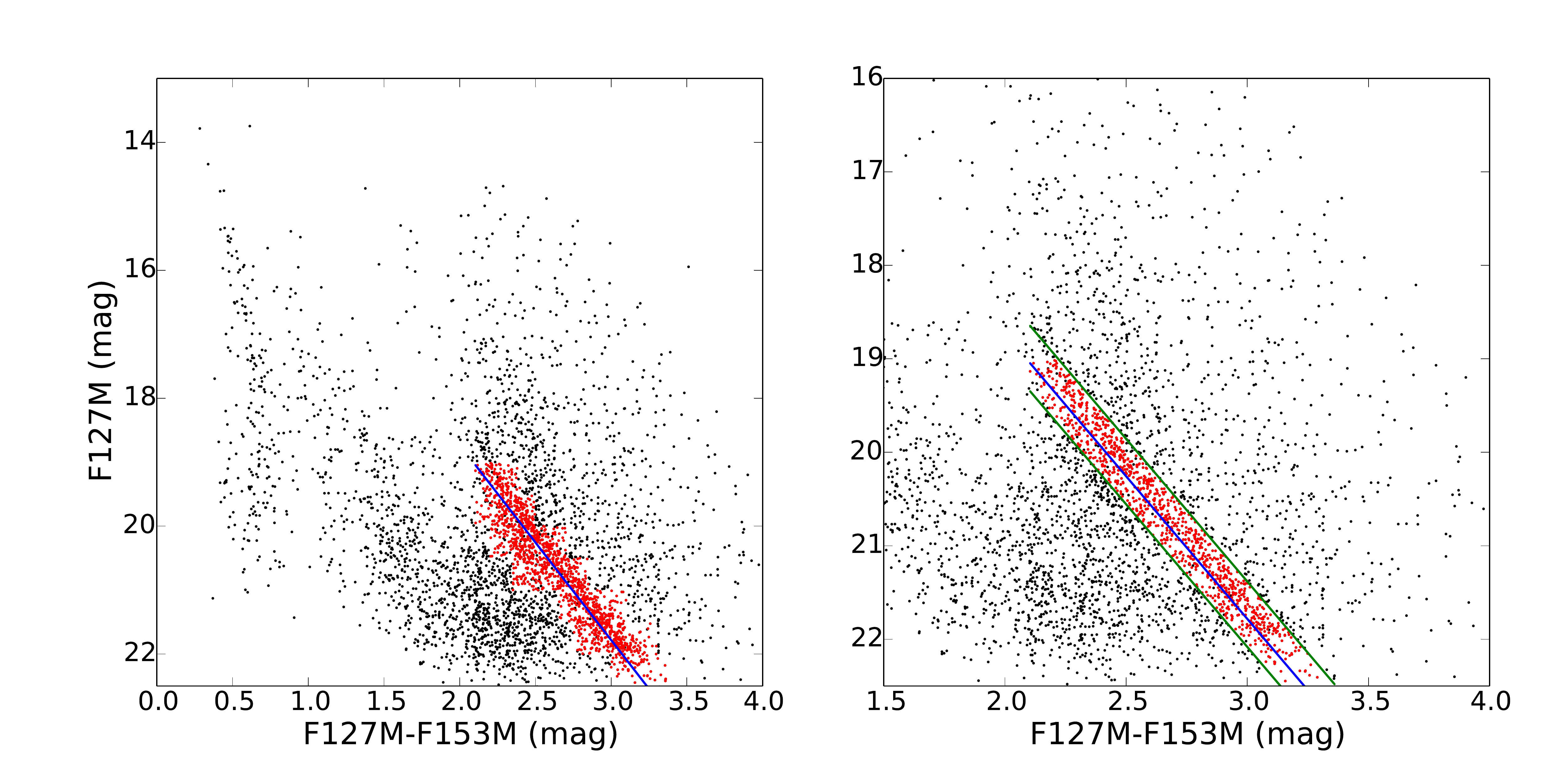}
\caption{The identification of Red Clump (RC) stars in the F127M vs. F127M - F153M color-magnitude diagram, as described in $\mathsection$ \ref{sec:extmap}. \emph{Left}: The full field CMD, with the stars used in the initial fit of the RC reddening vector in red and the reddening vector itself in blue. \emph{Right}: A zoomed-in view of the RC population, with the RC reddening vector and final identification criterion shown with blue and green lines, respectively. Stars falling between the green lines (red points) are identified as RC stars and are used to make the extinction map.}
\label{RC_ID}
\end{center}
\end{figure*}

\begin{figure}
\begin{center}
\includegraphics[scale=0.5]{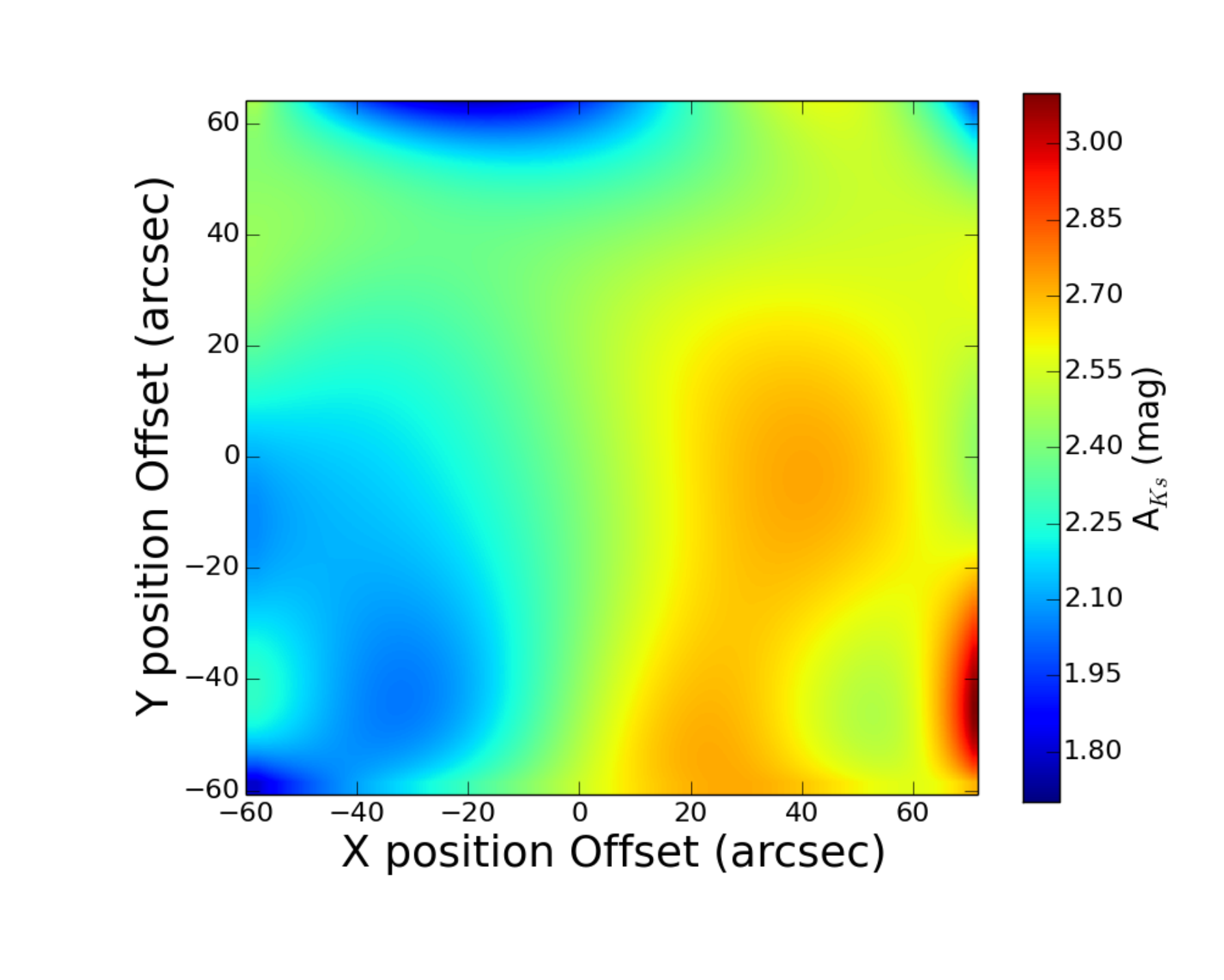}
\caption{The extinction map created via the spatial interpolation of the Red Clump extinction values as described in the text. The IR reddening law of  \citet{Nishiyama:2009fc} is used to calculate A$_{\lambda}$ at different wavelengths. Positions are given with respect to the cluster center located at (x, y) = (0, 0), and the axes are oriented same manner as Figure \ref{Arches_color}. Strong cluster candidates (P$_{member} >$ 0.7) have an average reddening of A$_{Ks}$ = 2.4 mag.}
\label{RedMap}
\end{center}
\end{figure}

Bulge RC stars are readily identified as a narrow population spread along the reddening vector in the F127M vs. F127M - F153M color-magnitude diagram (Figure \ref{RC_ID}). The spread of this population is primarily caused by differential reddening, which smears out what would normally be a tight clump of stars. In order to isolate RC stars, we use a PHOENIX model atmosphere \citep{A11} with typical RC parameters at solar metallicity \citep[T$_{eff}$ = 4700 K, log~\emph{g} = 2.40;][]{Mishenina:2006ij} to calculate the F127 vs. F127M - F153M reddening vector using the GC extinction law of \citet{Nishiyama:2009fc}. Keeping the slope of the reddening vector fixed, the y-intercept is fit to stars which fall within a broad area in the CMD around the RC population. We identify RC stars as those within a rectangle with the long axis centered on the reddening vector with the length of the short axis defined by the least crowded section of the RC bar (F127M - F153M $\approx$ 2.7). This corresponds to a width of constant value $\Delta$F127M = 0.7 mag. Identified cluster members ($\mathsection$ \ref{sec:Membership}) are removed from this sample. The extinction of each identified RC star is taken from the nearest point on the reddening vector in color-magnitude space.

We measure the extinction for 1027 RC stars identified across the field. These values are spatially interpolated using a 5th order bivariate spline to map the extinction at every position (Figure \ref{RedMap}). The typical error is $\sigma_{A_{Ks}}$ = 0.10 mags, as derived in $\mathsection$\ref{sec:Membership}. Extinction values range from 1.8 $<$ A$_{Ks}$ $<$ 3.0 with a median of A$_{Ks}$ = 2.4 for cluster members. This range is in agreement with the reddening map of \citet{Habibi:2013th}, who find 1.6 $<$ A$_{Ks}$ $<$ 3.3 also using a \citet{Nishiyama:2009fc} reddening law.

\subsection{Completeness Analysis}
\label{sec:complete}
In order to accurately measure the radial density profile of the Arches Cluster we must conduct an extensive completeness analysis on our astrometry pipeline. In addition to the sensitivity threshold of our observations, stars may be missed due to source confusion and proximity to bright and/or saturated stars. These effects are especially relevant for the dense central region of the cluster. To quantify our completeness we perform an artificial star injection and recovery test, planting 400,000 stars in each image and determining which are recovered to sufficient accuracy and precision as a function of spatial position and magnitude. The magnitudes of the artificial stars are drawn from the observed CMD of the field in order to best simulate the photometric properties of the observed stars. These magnitudes are then perturbed by a random amount reflecting the photometric uncertainty (assumed to be Gaussian distributed) of the real star they are simulating. The same set of artificial stars is applied to all observations. This analysis assumes that the artificial star measurement errors match those of the observed stars, which we test in Appendix \ref{sec:Artstar_appendix}.

The conditions an artificial star must fulfill in order to be considered as recovered matches the criteria applied to the real data. Within a given epoch, a recovered artificial star must: 1) be detected in at least 75\% of the images within that epoch; 2) have position and magnitude errors less than 1.5 mas (required for a proper motion precision better than or equal to 0.65 mas yr$^{-1}$) and 0.06 mag, respectively; and 3) have a measured position and magnitude within 0.5 pix (60 mas) and 0.5 mag of the planted values to guard against misidentification. In addition, artificial stars must be recovered in all 3 F153M epochs, which is required of the observed stars in order to derive their proper motions ($\mathsection$ \ref{sec:obs}). After detection/non-detection, the extinction map is used to differentially de-redden the artificial stars to the mean extinction of the cluster (A$_{Ks}$ = 2.4 mag). The fraction of recovered artificial stars to the total number of planted artificial stars represents the completeness fraction as a function of position and differentially de-reddened magnitude.

\begin{figure*}
\begin{center}
\includegraphics[scale=0.45]{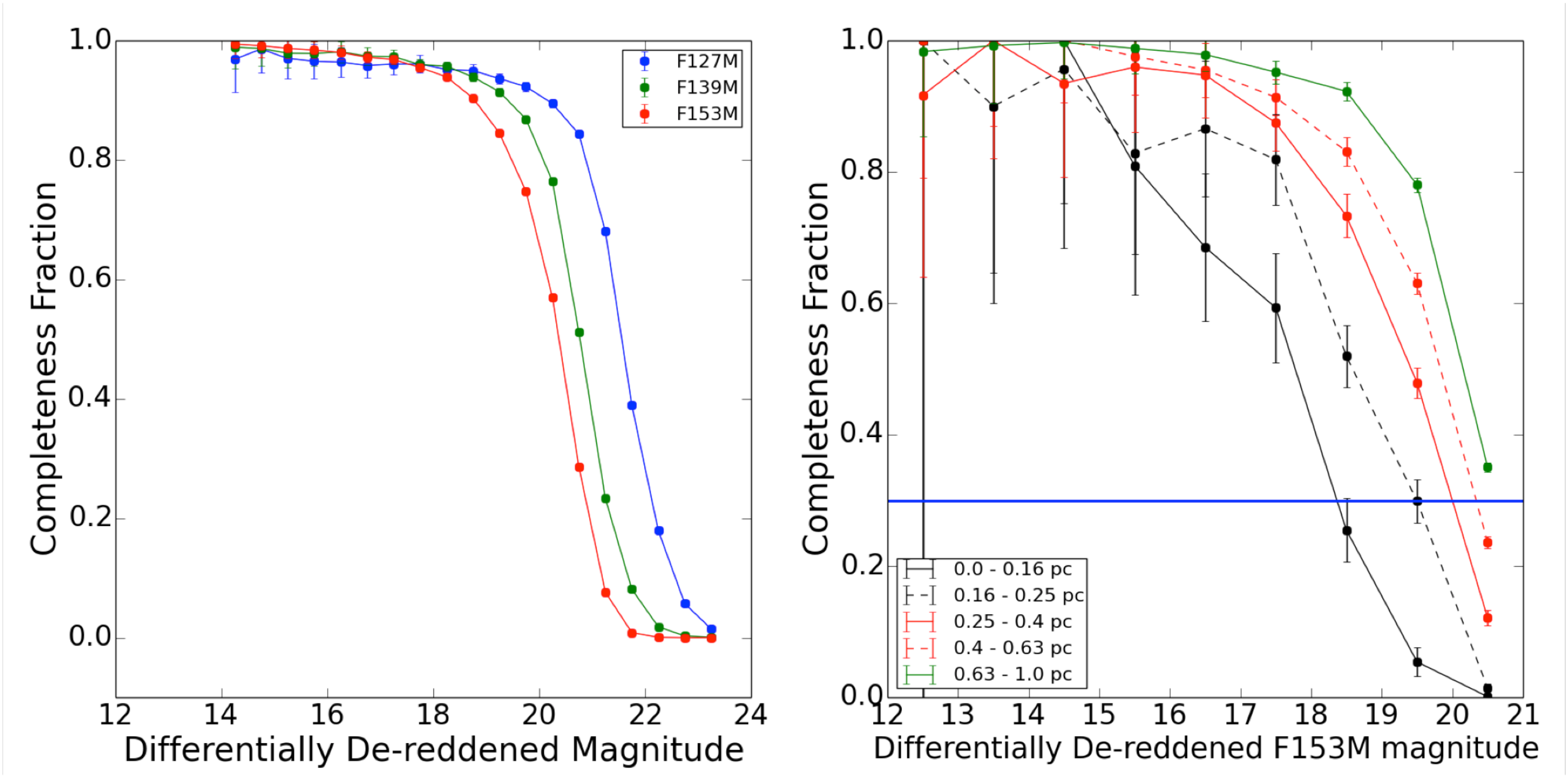}
\caption{Completeness as a function of observed magnitude and radius. \emph{Left}: Completeness as a function of differentially de-reddened magnitude (A$_{Ks}$ = 2.4 mag) for the F127M, F139M, and F153M filters. These completeness values are calculated over the entire field. \emph{Right}: The F153M completeness as a function of differentially de-reddened magnitude in different radius bins. The blue line marks a minimum completeness of 30\%, which is achieved down to F153M $=$ 20 mag for R $>$ 0.25 pc. This sets the faint-end magnitude limit and inner radius limit for the radial profile.}
\label{comp_mag_radius}
\end{center}
\end{figure*}

The resulting completeness curves as a function of differentially de-reddened magnitude and of radius are presented in Figure \ref{comp_mag_radius}. Over the full field we achieve greater than 50\% completeness down to F153M $=$ 20 mag. However, the completeness in the inner 6.25" ($\sim$0.25 pc) of the cluster is significantly lower due to stellar crowding, falling to 30\% by F153M $=$ 18.5 mag. As a result, we restrict the following analysis to observed stars with R $>$ 6.25" and differentially de-reddened magnitudes brighter than F153M = 20 mag.

\section{Results}
\subsection{Cluster Membership}
\label{sec:Membership}
\begin{figure}
\begin{center}
\includegraphics[scale=0.35]{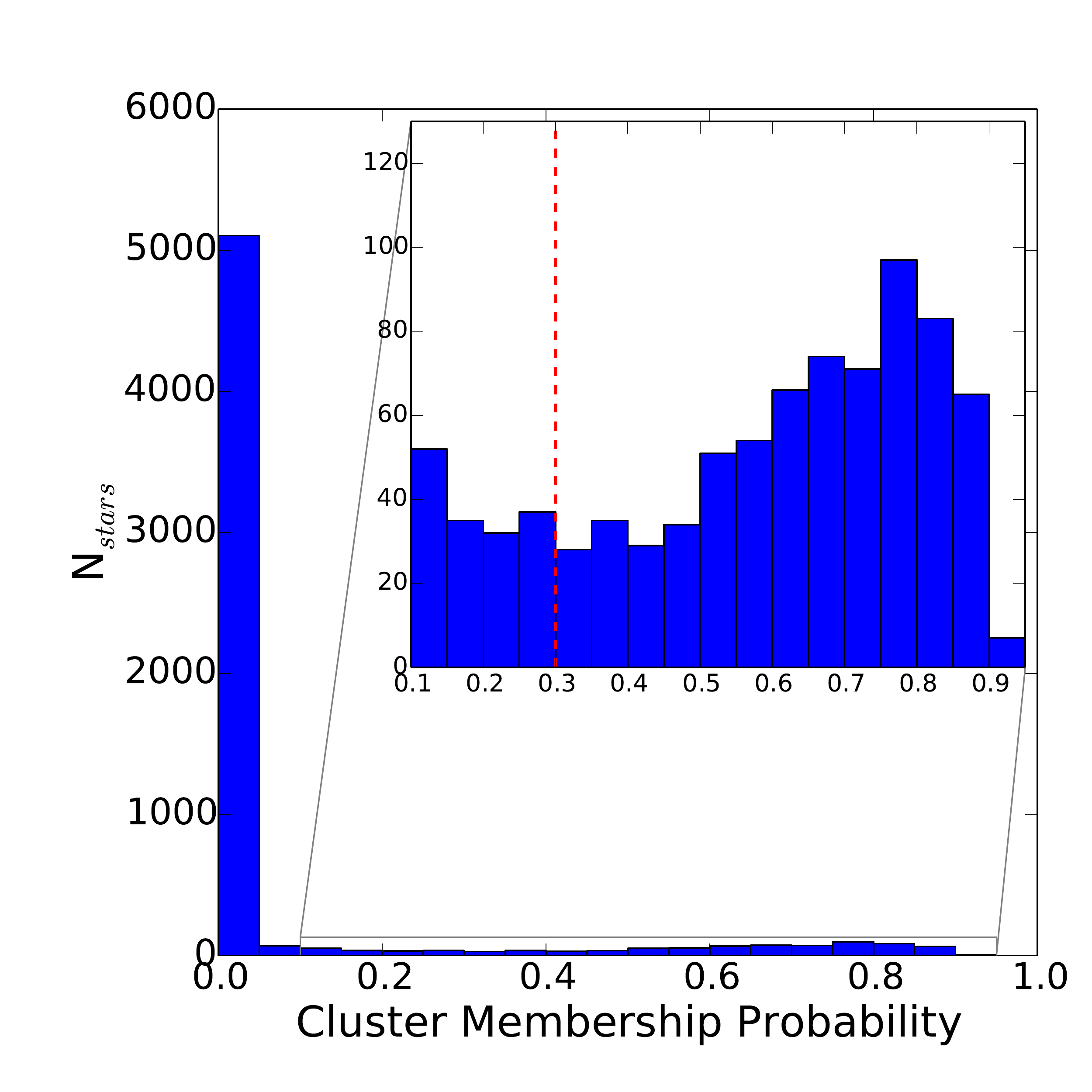}
\caption{A histogram of the membership probabilities obtained for the sample. All objects with P$_{cluster} \ge$ 0.3 (red line, inset plot) are considered in the profile analysis, weighted by their membership probability. Of $\sim$6000 stars examined, 701 meet this criterion, with membership probabilities that sum to 446.8.}
\label{prob_hist}
\end{center}
\end{figure}

\begin{figure*}
\begin{center}
\includegraphics[scale=0.7]{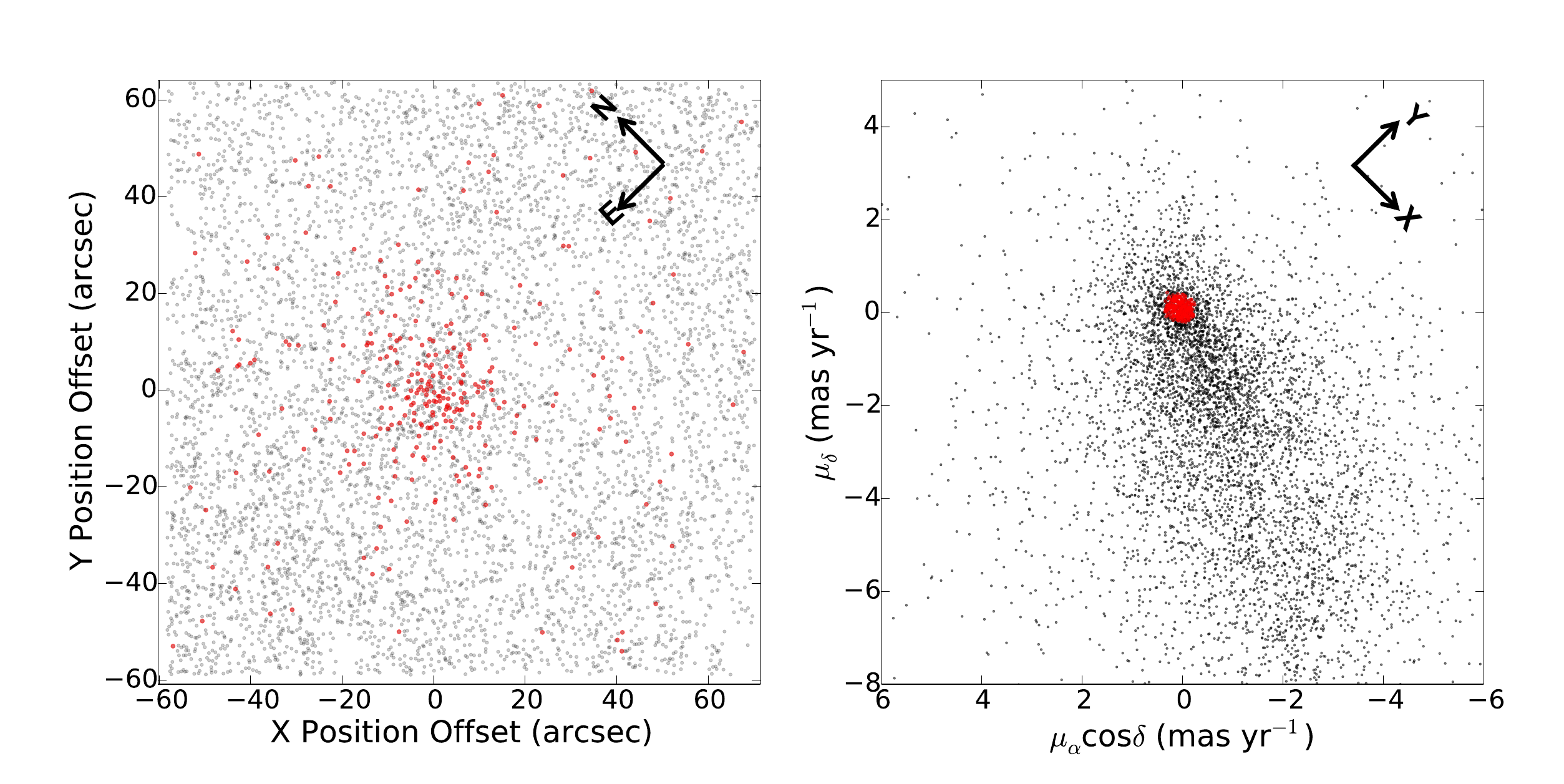}
\caption{The spatial (\emph{left}) and kinematic (\emph{right}) positions of strong cluster candidates (P$_{member} >$ 0.7), marked as red points, compared to the rest of the sample in black points. The spatial positions are plotted in arcseconds relative to the cluster center and are in image coordinates (same orientation as Figure \ref{Arches_color}). Proper motions are plotted in projected equatorial coordinates. The radial profile of the Arches includes all stars with P$_{member} >$ 0.3, a larger sample than is shown here.}
\label{members}
\end{center}
\end{figure*}

With the kinematic properties of the cluster and field populations determined ($\mathsection$ \ref{sec:pops}), we calculate the probability of cluster membership for each star based on its proper motion:

\begin{equation}
P_{member}^i = \frac{\pi_c P_{c}^i}{\pi_c P_{c}^i + \sum_k^K{\pi_k P_{k}^i}}
\end{equation}

\noindent where $\pi_c$ and $\pi_k$ are the fraction of total stars in the cluster and $k$th field Gaussian, respectively, and P$_{c}^i$ and P$_{k}^i$ are the probability of $i$th star being part of the cluster and $k$th field Gaussian, respectively. A histogram of the resulting cluster membership probabilities is shown in Figure \ref{prob_hist}. In the following analysis we include all stars with P$_{member} >$ 0.3, weighted by their individual membership probabilities. This criteria selects 701 stars which represent 446.8 ``cluster members'' based on the sum of the cluster membership probabilities. We consider stars with P$_{member} > 0.7$ as strong cluster candidates, whose distributions in position and velocity space are shown in Figure \ref{members}.

\begin{figure*}
\begin{center}
\includegraphics[scale=0.7]{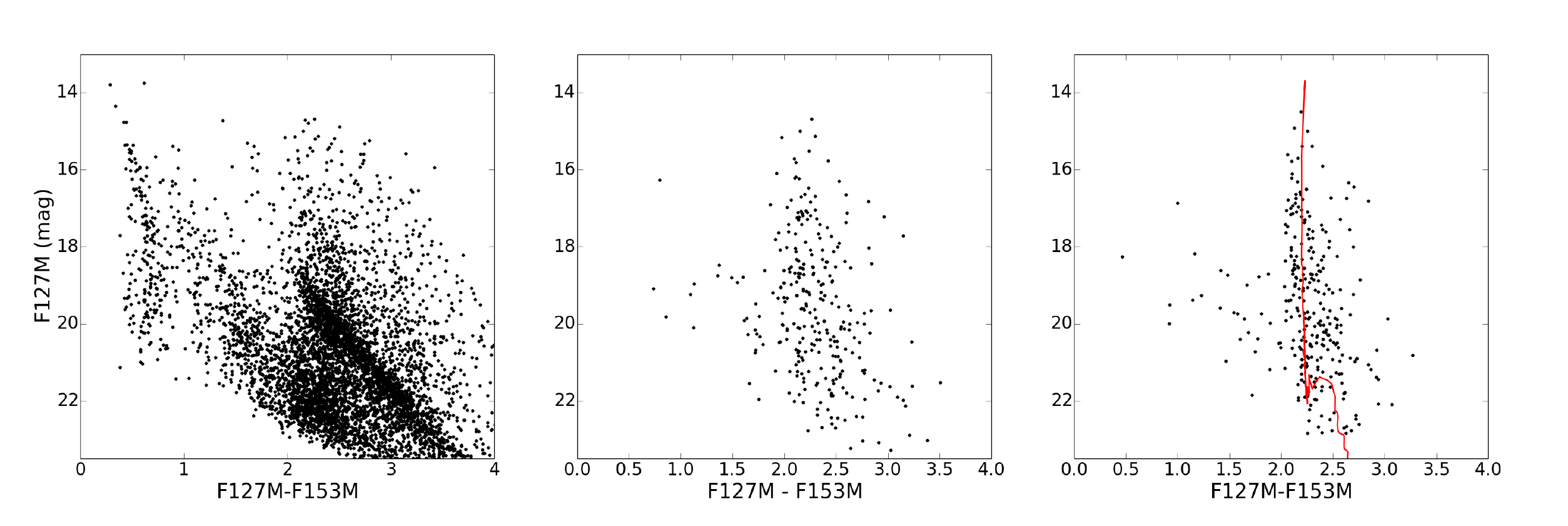}
\caption{The F127M vs. F127M - F153M color-magnitude diagram for the full sample (\emph{left}), strong cluster candidates (P$_{member} >$ 0.7, \emph{middle}), and strong cluster candidates after being differentially de-reddened using the extinction map (\emph{right}). The reddening correction noticeably tightens the color dispersion and blue edge of the population. The differentially de-reddened CMD is consistent with a theoretical 2.5 Myr cluster isochrone at 8000 pc with A$_{Ks}$ = 2.4 mag, overlaid in red. The isochrone is created using the pre-main sequence evolutionary models of \citet{Siess:2000fv} for M $<$ 7 M$_{\odot}$ and the main sequence models with rotation of \citet{Meynet:2003ty} for M $>$ 9 M$_{\odot}$, with an interpolation between the models over the missing mass range. }
\label{clusterCMD}
\end{center}
\end{figure*}

By applying the extinction map derived in $\mathsection$\ref{sec:extmap} to the strong cluster candidates we create a differentially de-reddened F127M vs F127M - F153M CMD of the cluster (Figure \ref{clusterCMD}). The improvement relative to the uncorrected CMD is noticeable in both the overall color dispersion and definition of the blue edge. Stars with colors more blue than the blue edge are very likely field contaminants, while the the scatter along the redward edge of the cluster may be caused by intrinsic reddening of the objects themselves, perhaps due to circumstellar disks \citep{Stolte:2010fv}.

The remaining color dispersion of the differentially de-reddened CMD provides an estimate of the uncertainties of the extinction map. Between 16 $<$ F127M $<$ 21 mag the median color dispersion is 0.36 mags. Assuming that the photometric uncertainties in each filter are negligible and adopting the \citet{Nishiyama:2009fc} extinction law, this color dispersion corresponds to an extinction error of $\sigma_{A_{F153M}}$ = 0.18 mags ($\sigma_{A{Ks}}$ = 0.10 mags).

The mean extinction of the strong cluster candidates is A$_{Ks}$ = 2.42 $\pm$ 0.14 mag. The inner region of the cluster (R $<$ 0.4 pc) exhibits a tight range of reddening values from 2.33 $<$ A$_{Ks}$ $<$ 2.53 mag, while the outer region of the cluster (R $>$ 0.4 pc) exhibits a much wider range from 2.04 $<$ A$_{Ks}$ $<$ 2.76 mag. In the literature, there are variations in the measured extinction for the Arches due to different methodologies and reddening laws. Using the same \citet{Nishiyama:2009fc} extinction law with the CMD sliding method, \citet{Habibi:2013th} find A$_{Ks}$ = 2.6 $\pm$ 0.2 mag for 0.2 pc $<$ R $<$ 0.4 pc and A$_{Ks}$ = 2.6 $\pm$ 0.3 mag for 0.4 pc$<$ R $<$ 1.5 pc. Also using the CMD sliding method, \citet{Espinoza:2009bs} and \citet{Kim:2006fy} find higher values of A$_{Ks}$ = 2.97 and 3.1 for R $<$ 0.4 pc using the extinction laws of \citet{Fitzpatrick:2004ud} and \citet{Rieke:1989kl}, respectively. Our measurements are consistent with previous measurements made using the same extinction law.

\subsection{Radial Density Profile}
\label{sec:profile}
Using stars with P$_{member} >$ 0.3, we construct the radial profile of the Arches Cluster using the Bayesian methodology described by \citet{Do:2013il}. This allows us to construct an un-binned profile that simultaneously incorporates cluster membership probabilities, image completeness, and geometric area corrections at large radii to account for incomplete area coverage. As discussed in $\mathsection$\ref{sec:complete}, only stars with differentially de-reddened magnitudes of F153M $\leq$ 20 mag and R $>$ 6.25" (0.25 pc) are considered. We adopt a single power law as our likelihood function:

\begin{equation}
\label{eq:PL}
L_i(r, \Gamma, b) = A_0r_i^{-\Gamma} + b
\end{equation}

\noindent where $r_i$ is the radius of the $i$th star and the field contamination $b$ is assumed to be constant across the image. The profile amplitude $A_0$ is calculated such that the integral of the radial profile yields the total number of cluster members observed after membership probability, completeness, and area corrections.

\begin{figure*}
\begin{center}
\includegraphics[scale=0.3]{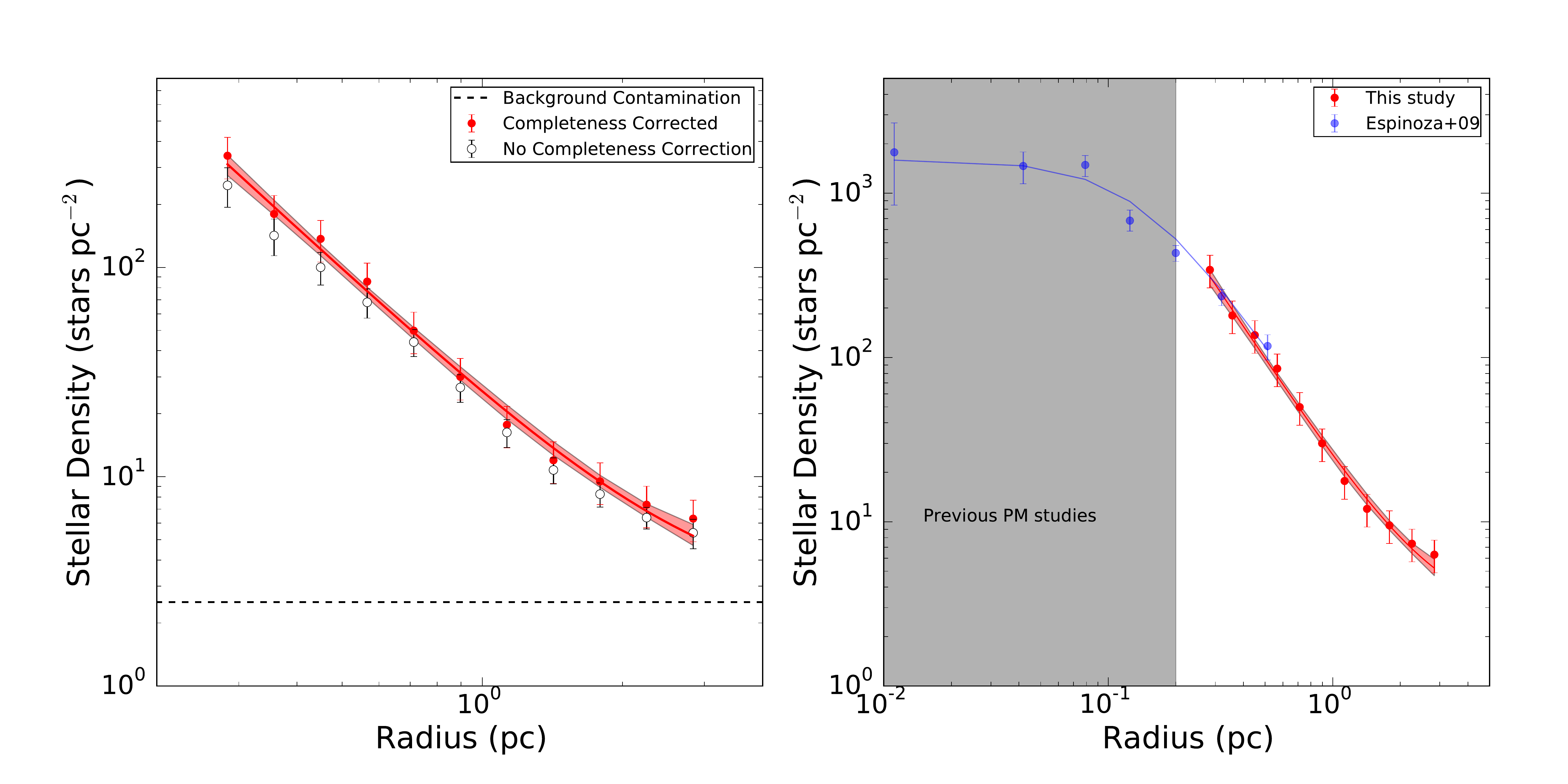}
\caption{The radial profile of the Arches Cluster. \emph{Left}: The best-fit power law model as described in the text and Table \ref{PL_results}. The red line represents the power-law fit to the unbinned data and the black dotted line the residual field contamination value. A binned profile is included to guide the eye; the black open points are the binned profile before completeness correction and the red solid points the binned profile after completeness correction. Uncertainty in the fit (1$\sigma$) is captured by the red shaded region, which spans the standard deviation of 1000 profiles randomly drawn from the joint posterior distribution. Note that the binned profile is presented only for comparison and does not affect the fit. \emph{Right}: Same as left, but with the inner cluster profile from \citet{Espinoza:2009bs} added in blue (10 M$_{\odot}$ $<$ M $<$ 120 M$_{\odot}$) and scaled to our profile. The radius range probed the previous proper motion study of \citet{Clarkson:2012ty} is shaded in grey.}
\label{ProfilePL}
\end{center}
\end{figure*}

The total likelihood $\mathcal{L}$ is then the product of the individual likelihoods for $N$ total stars:

\begin{equation}
\label{eq:PL_likelihood}
\mathcal{L}  = \sum_i^N w_i(r)\log L_{i}(r, \Gamma, b)
\end{equation}
\begin{equation*}
w_i(r) = \frac{P_i}{A_i(r)C_i(r)}
\end{equation*}
\\

\noindent where $P_i$ is the membership probability of the $i$th star, $A_i$(r) is the relative fraction of observed area at the star's radius relative to an infinite field of view ($A_i = 1.0$ for 0 $<$ r $\leq$ 60", $A_i$ $<$ 1.0 for r $>$ 60"), and $C_i$(r) is the completeness at that star's radius. A summary of our best-fit model and subsequent results are presented in Table \ref{PL_results} and Figure \ref{ProfilePL}.  A binned profile is included for comparison, with errors calculated from the Poisson uncertainties in the completeness correction and observed profile as well as the uncertainties in the extinction map. These are captured by recalculating the stellar density in each radius bin using a magnitude cut brighter or fainter than F153M = 20 mag by the map error value ($\sigma_{A_{F153M}}$ = 0.18 mag). The half-light radius of the profile is 0.48 pc, largely consistent with previous studies (0.4 pc, \citeauthor{Figer:1999px} 1999). The bivariate posterior distributions for these parameters are presented in Appendix \ref{sec:posteriors_appendix}.

Included in the right panel of Figure \ref{ProfilePL} is the radial profile for the inner part of the Arches from \citet{Espinoza:2009bs}, which spans a stellar mass range of 10 M$_{\odot}$ $<$ M $<$ 120 M$_{\odot}$ out to R = 0.4 pc. There is good agreement between the shape of the two profiles, though the absolute values of the \citet{Espinoza:2009bs} profile must be scaled. This is necessary due to differences in sensitivity and treatment of cluster membership. We note that our profile spans far beyond the limits of previous astrometric studies of the Arches cluster, which are restricted to R $<$ 0.2 pc. We leave the combination of these astrometric data sets and the presented data set to a future paper.

\begin{deluxetable*}{l l c c c c}
\tablewidth{0pt}
\tabletypesize{\footnotesize}
\tablecaption{Power-Law Profile Fit Results}
\tablehead{
\colhead{} & \colhead{Bin} & \colhead{N$_{stars}$\tablenotemark{a}} & \colhead{Power-law slope} & \colhead{Field Contamination} & \colhead{Normalization Constant} \\
\colhead{} & \colhead{} & \colhead{} & \colhead{$\Gamma$} & \colhead{$b$ (stars / pc$^2$)} & \colhead{$A_0$ (stars / pc$^2$)}
}
\startdata
Full Cluster &  & 451.0 & 2.06 $\pm$ 0.17 &  2.52 $\pm$ 1.32 & 23.09 $\pm$ 3.5  \\
Split by Mass (2-bin) & High Mass\tablenotemark{b} & 106.5 & 2.70 $\pm$ 0.35 & 0.64 $\pm$ 0.48 & 3.50 $\pm$ 1.22 \\
& Low Mass\tablenotemark{c} & 354.5 & 1.75 $\pm$ 0.15 & 0.78 $\pm$ 1.37 & 20.33 $\pm$ 2.75  \\

Split by Mass (3-bin) & High Mass\tablenotemark{d} & 129.6 & 2.75 $\pm$ 0.37 & 1.47 $\pm$ 0.48 & 3.36 $\pm$ 1.26 \\
& Intermediate Mass\tablenotemark{e} & 163.6 & 2.00 $\pm$ 0.28 & 2.04 $\pm$ 0.56 & 6.38 $\pm$ 1.46  \\
& Low Mass\tablenotemark{f} & 165.7 & 2.29 $\pm$ 0.30 & 2.09 $\pm$ 0.53 & 6.47 $\pm$ 1.48  \\

Split by Direction\tablenotemark{g}: & Parallel  & 225.7 & 1.86 $\pm$ 0.17  & 2.32 $\pm$ 0.88 & 22.57 $\pm$ 1.9  \\
& Perpendicular & 226.0 & 2.19 $\pm$ 0.18 & 2.29 $\pm$ 0.82 & 21.35 $\pm$ 2.43  \\
Prior\tablenotemark{h} &  &  & U(0.5, 4.5) & U(0,8), G(2.52,1.32)\tablenotemark{i} & ---
\enddata
\label{PL_results}
\tablenotetext{a}{Weighted by membership probability and corrected for completeness.}
\tablenotetext{b}{F153M $<$ 17 mag (M $>$ $\sim$13 M$_{\odot}$)}
\tablenotetext{c}{17 $<$ F153M $<$ 20.0 mag ($\sim$2.5 M$_{\odot} <$ M $<$ $\sim$13 M$_{\odot}$)}
\tablenotetext{d}{F153M $<$ 17.3 mag (M $>$ $\sim$12 M$_{\odot}$)}
\tablenotetext{e}{17.3 $<$ F153M $<$ 18.8 mag ($\sim$6 M$_{\odot} <$ M $<$ $\sim$12 M$_{\odot}$)}
\tablenotetext{f}{18.8 $<$ F153M $<$ 20.0 mag  ($\sim$2.5 M$_{\odot} <$ M $<$ $\sim$6 M$_{\odot}$)}
\tablenotetext{g}{Relative to the direction of the Arches cluster orbit}
\tablenotetext{h}{Uniform distributions: U(min, max), where min and max are bounds of the distribution; Gaussian distributions: G($\mu, \sigma$), where $\mu$ is the mean and $\sigma$ is the standard deviation}
\tablenotetext{i}{Adopted U(0,8) for the full cluster profile fit, G(2.52, 1.32) for the directional profile fit}
\end{deluxetable*}

To quantitatively assess whether the power-law model is an appropriate one for the observed profile, we conduct a posterior predictive analysis using $\chi^2$ as the test statistic \citep{Gelman_book}. We randomly select 1000 sets of model parameters from the joint posterior distribution and generate artificial binned profiles from these models. Each data point within the artificial profiles is shifted by an offset randomly drawn from a normal distribution with a width equal to the uncertainty in that value, determined from the combination of the poisson uncertainty and the uncertainty in the completeness correction. We then calculate a $\chi^2$ value for each binned profile with respect to the best-fit model to the observations:

\begin{equation}
\chi^{2}(\Gamma, b)  = \sum_{j=1}^{n} \frac{(P_j^{bin} - P_j^{model} )^2}{\sigma^2_j}
\label{chieq}
\end{equation}
\\
where $P_j^{bin}$ is the $j$th point in the binned profile with uncertainty $\sigma_j$ and $P_j^{model}$ is the value predicted for the $j$th bin by the best-fit model. We find that only 1\% of these $\chi^2$ values are lower than the $\chi^2$ value for the observed binned profile and conclude that a power-law model is a good fit to the data. We emphasize that this $\chi^2$ statistic and binned profiles are used to test the validity of the model, not to fit the model itself.

Previous studies of other YMCs have shown that these objects often appear to have extended radial profiles without signs of tidal truncation \citep{Elson:1987qe,Mackey:2003eu,Mackey:2003oq,McLaughlin:2005rr}. These studies fit the profiles using a model defined by \citet{Elson:1987qe}, hereafter referred to as an EFF87 profile model. We fit our profile with this model, adding a constant term $b$ for field contamination:

\begin{equation}
\label{Elson}
\Sigma(r) = \Sigma_0 \left( 1 + \frac{r^2}{a^2} \right)^{-\gamma / 2} + b
\end{equation}
\\
where $a$ is related to the core radius $r_c$ of the cluster:

\begin{equation}
r_c = a \left( 2^{2 / \gamma} - 1 \right)^{1/2}
\end{equation}

We use the same Bayesian framework as described above, only with this profile as the likelihood function. Since incompleteness prevents our profile from stretching into the core region of the Arches we adopt the core radius $r_c$ determined by \citet{Espinoza:2009bs} of 0.14 $\pm$ 0.05 pc in our model. The consequent fit is nearly identical to the single power law model fit, with parameter values provided in Table \ref{ElsonModel}. We obtain $\gamma$ = 2.3 $\pm$ 0.2, which is consistent with the range and median values of 2.01 -- 3.79 and 2.59 determined from a sample of LMC and SMC YMCs by \citet{Mackey:2003eu,Mackey:2003oq}. Given that we have no information about $r_c$ from our profile and the marginal difference between this profile and the power-law fit, we proceed with the single power-law model.

\begin{deluxetable*}{l l c}
\tablewidth{0pt}
\tabletypesize{\footnotesize}
\tablecaption{EFF87 Profile Fit Results}
\tablehead{
\colhead{Parameter\tablenotemark{a}} & \colhead{Prior\tablenotemark{b}} & \colhead{Result}
}
\startdata
$\gamma$ & U(0.5, 4.5) & 2.3 $\pm$ 0.2 \\
$b$ (stars pc$^{-2}$) & U(0, 8) & 3.3 $\pm$ 1.2  \\
$a$ (pc) &  --- & 0.13 $\pm$ 0.03  \\
$\Sigma_0$ (stars pc$^{-2}$) &  --- & 2209 $\pm$ 929
\enddata
\label{ElsonModel}
\tablenotetext{a}{Description of parameters: $\gamma$ = Outer power-law slope; $b$ = field contamination, $a$ = core radius, $\Sigma_0$ = Normalization factor}
\tablenotetext{b}{Uniform distributions presented as U(min, max), where min and max are bounds of the distribution}
\end{deluxetable*}

\subsubsection{Mass Segregation}
Evidence for mass segregation in the Arches cluster has been found in the flattening of the mass function toward the cluster center \citep{Figer:1999lo, Stolte:2002zr, Espinoza:2009bs, Habibi:2013th} and a shallower radial profile for stars between $\sim$10 - 30 M$_{\odot}$ compared to stars between $\sim$30 - 120 M$_{\odot}$ \citep{Espinoza:2009bs}. However, in addition to being dependent on photometric cluster membership, these results rely on measurements in the dense innermost regions of the cluster (R $<$ 15") where completeness is lowest for low-mass stars due to stellar crowding \citep{Ascenso:2009rt}. We avoid this inner region and instead examine mass segregation in the less-dense outer regions of the cluster.

\begin{figure*}
\begin{center}
\includegraphics[scale=0.5]{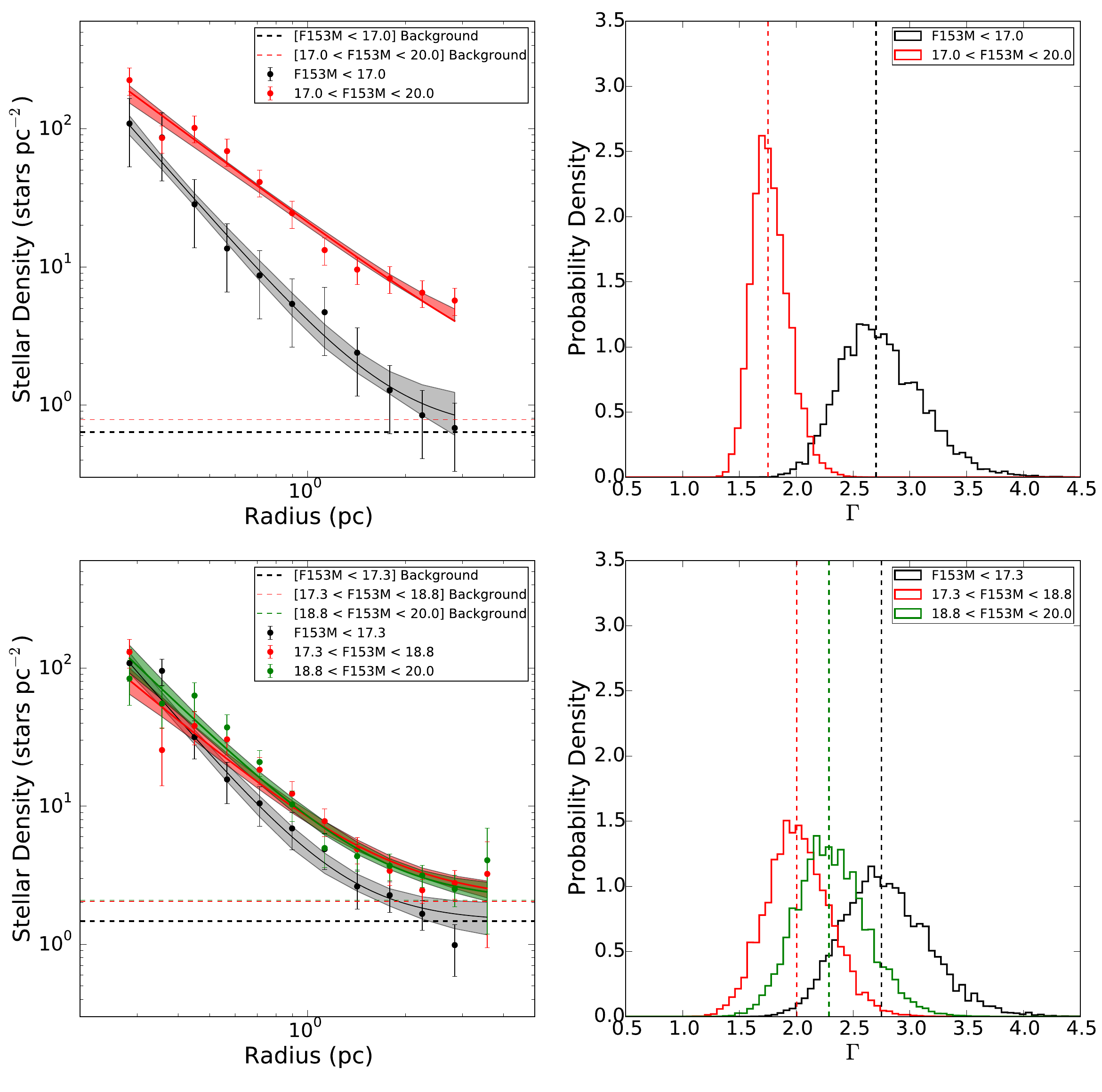}
\caption{Mass segregation in the Arches cluster. \emph{Top:} The clearest evidence for mass segregation is found in the steepening of the profile for high-mass stars (black; F153M $\leq$17 mag, or M $\geq$$\sim$13 M$_{\odot}$) compared to low-mass stars (red; F153M $>$17 mag, or M $<$$\sim$13 M$_{\odot}$). On the \emph{left}, the solid lines show the power law profile fit to the unbinned data, the dotted lines show the residual field contamination values, and the data points show the binned profiles after completeness correction. 1$\sigma$ model uncertainties are shown as the shaded regions. The posterior distributions of the power-law slope $\Gamma$ are shown to the \emph{right}. The slopes differ by 2.5$\sigma$, and a Kolmogorov-Smirnov test rejects the hypothesis that the profiles are drawn from the same parent distribution. \emph{Bottom:} Similar to above, but splitting the sample into three subsamples by magnitude such that each magnitude bin contains $\sim$130 stars before completeness correction. Mass segregation remains evident in the steepening of the brightest profile compared to the other two profiles. The best parameter values for all fits are presented in Table \ref{PL_results}. }
\label{ProfileMS}
\end{center}
\end{figure*}

Following \citet{Espinoza:2009bs}, we separate our radial profile as a function of differentially de-reddened magnitude to test for mass segregation. These magnitudes are a good proxy for stellar mass, as only a small fraction of stars in the Arches ($\sim$6\%) have been found to exhibit IR~excess emission from circumstellar disks which could bias the photometry \citep{Stolte:2010fv}. Adopting the single power-law model described above (Equations \ref{eq:PL}, \ref{eq:PL_likelihood}), we find the power-law slope of stars brighter than F153M = 17 mag (M $>$$\sim$13 M$_{\odot}$) to be notably steeper than the slope of stars between F153M = 17 -- 20 mag ($\sim$2.5 M$_{\odot}$$<$ M $<$$\sim$13 M$_{\odot}$). A Kolmogorov-Smirnov test finds the probability of these profiles being drawn from the same parent distribution to be $<$0.05\%, demonstrating that mass segregation is present throughout the spatial extent of the cluster. The profiles in these different magnitude bins are shown in Figure \ref{ProfileMS}, with the fit summarized in Table \ref{PL_results} and accompanying bivariate posterior distributions in Appendix \ref{sec:posteriors_appendix}.

The adopted magnitude separation is an optimization between obtaining a large enough sample for good statistics in the bright-star profile and showing the mass segregation, which become less evident with fainter magnitude cuts. To demonstrate this, we split the sample into three subsets by magnitude such that each magnitude bin contains $\sim$130 cluster members before completeness corrections in Figure \ref{ProfileMS}. These magnitude bins correspond to F153M $<$ 17.3 mag (M $>$$\sim$12 M$_{\odot}$), F153M = 17.3 -- 18.8 mag ($\sim$6 M$_{\odot}$$<$ M $<$$\sim$12 M$_{\odot}$), and F153M = 18.8 -- 20 mag ($\sim$2.5 M$_{\odot}$$<$ M $<$$\sim$6 M$_{\odot}$). Mass segregation remains evident with the brightest (thus highest mass) profile being noticeably steeper than the other two profiles, while the intermediate and faint-star profiles are more similar to one another.

We caution that the conversion from observed magnitude to mass is highly uncertain due to uncertainties in evolutionary models, especially in the pre-main sequence. To determine the stellar masses at the magnitudes presented above, we adopt a cluster isochrone with the nominal properties of the Arches cluster (age = 2.5 Myr, distance = 8000 pc, A$_{Ks}$ = 2.4 mag) constructed using a combination of Geneva models with initial rotation speed of 300 km s$^{-1}$ \citep{Meynet:2003ty} and \citet{Siess:2000fv} pre-main sequence models, as discussed in \citet{Lu:2013wo}. However, a more accurate conversion from magnitude to mass (and a more detailed examination of cluster mass segregation) will be the focus of a future paper.

\subsubsection{The Search for Tidal Tails}
\label{sub:TT}
Given the strong gravitational fields near the GC, the Arches cluster is expected to have tidal tails leading and trailing its orbit. Such structures have been observed for globular clusters and have yielded insight to the object's orbit and the gravitational potential of the Galaxy \citep[e.g.,][]{Odenkirchen:2001fr, Odenkirchen:2003jk, Grillmair:2006ff}. Adopting the model of the initial conditions of the Arches from \citet{Harfst:2010ys}, the 3D velocity from \citet{Clarkson:2012ty}, and assuming a current position 100 pc in front of the GC, \citet{Habibi:2014qf} predict that the Arches should have tidal tails extending 20 pc ($\sim$500") along the Galactic plane. To search for these structures we compare the radial profiles parallel and perpendicular to the cluster's bulk velocity, which is consistent with the direction of the Galactic plane \citep{Clarkson:2012ty}.

Tidal tails would cause an asymmetry in these profiles, either as a steepening or truncation of the perpendicular profile relative to the parallel profile as the cluster is stretched and sheared by the Galactic tidal field (i.e. Figure 5 of \citeauthor{Odenkirchen:2003jk} 2003). Using the single power-law model, we do not find evidence of a significant difference between the parallel and perpendicular profile slopes (Table \ref{PL_results}, Figure \ref{ProfileTT}). Bivariate posterior distributions for these profiles are are presented in Appendix \ref{sec:posteriors_appendix}. While the binned profiles might appear to be discrepant at $\sim$1 -- 1.5 pc, a Kolmogorov-Smirnov test of these profiles in the region of highest completeness (0.5 pc $<$ R $<$ 3.0 pc) concludes that the profiles are drawn from the same distribution with a probability of $\sim$16\%. Thus, we cannot conclude that profiles are statistically different. This conclusion does not change when different cluster membership probability cuts are adopted (P $>$ 0.7, for example). Further observations of the Arches cluster at large radii are needed to detect the presence of tidal tails.

\begin{figure*}
\begin{center}
\includegraphics[scale=0.3]{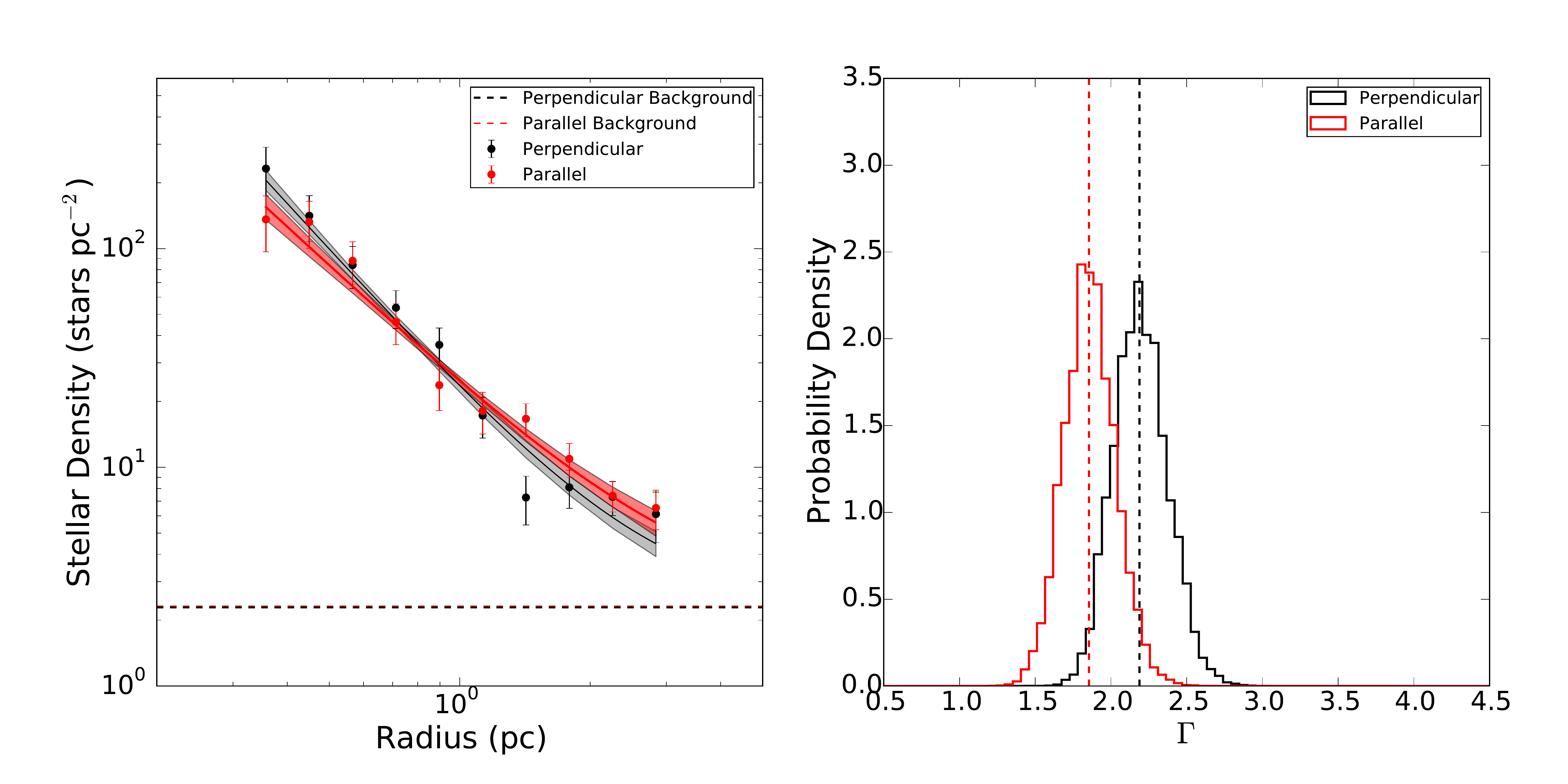}
\caption{The search for tidal tails in the Arches cluster. \emph{Left:} No significant asymmetries suggesting the presence of tidal tails are found in a comparison of the profile parallel (red) and perpendicular (black) to the cluster's orbit. Power-law fits, residual field contamination, and 1$\sigma$ uncertainties are shown in the same manner as Figure \ref{ProfilePL}, while the best-fit parameter values are presented in Table \ref{PL_results}. \emph{Right:} The posterior distributions for the power-law slope $\Gamma$ of the fitted profiles. The black and red dotted lines show the best-fit slopes for the perpendicular and parallel profiles, respectively, and which only differ by $\sim1.4\sigma$. A Kolmogorov-Smirnov test cannot reject the hypothesis that these profiles were drawn from the same parent distribution.}
\label{ProfileTT}
\end{center}
\end{figure*}

\subsection{Observed Tidal Radius}
\label{sec:tidal}
\begin{figure}
\begin{center}
\includegraphics[scale=0.35]{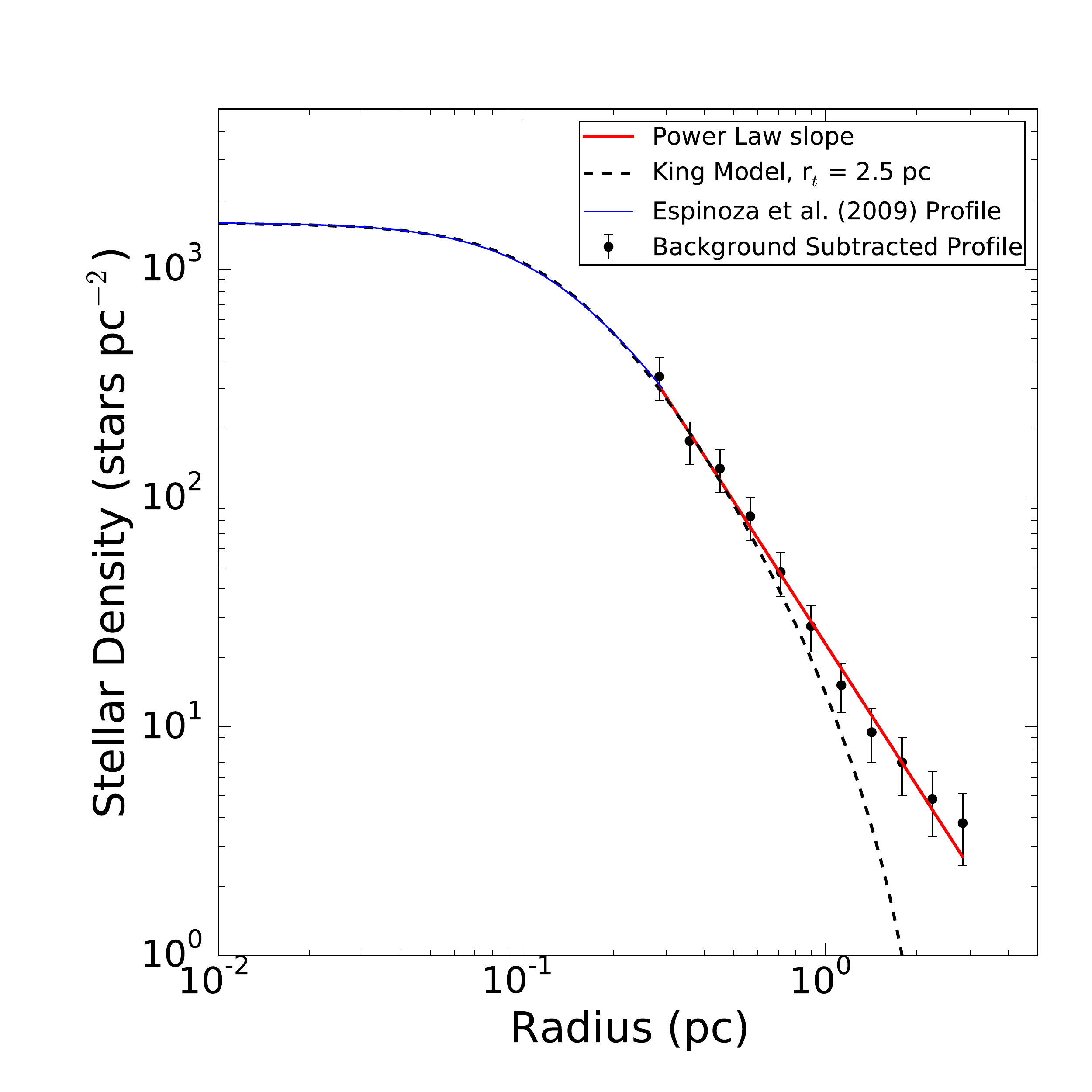}
\caption{The background-subtracted profile of the Arches cluster compared to a King profile with a tidal radius of r$_t$ = 2.5 pc. The Arches profile is composed of the best-fit power law slope (red line) and binned profile (black points) from this study, and the profile for the inner 0.4 pc measured by \citet{Espinoza:2009bs} scaled to our profile (blue line). The King profile model (black dotted line) is clearly discrepant with this profile at large radii. We place a 3$\sigma$ lower limit of 2.8 pc on the location of a King-like tidal radius in the Arches cluster.}
\label{King_fit}
\end{center}
\end{figure}

Utilizing the large field of view, we can directly constrain the spatial extent of the Arches cluster for the first time. This is a significant improvement over previous studies which were forced to estimate the tidal radius based on observations of the inner region of the cluster. For example, \citet{Kim:2000wd} compared the radial profile of massive stars in the Arches (M $>$ 20 M$_{\odot}$) out to 0.8 pc \citep{Figer:1999lo} to the radial profiles of similarly massive stars in N-body simulations of the cluster and found an expected tidal radius between 1 -- 1.2 pc. A second estimate by \citet{Portegies-Zwart:2002hc} placed the tidal radius at 1.6 -- 2.5 pc, based on a highly model-dependent analysis of the mass segregation observed in the same \citet{Figer:1999lo} profile. However, our study shows that the Arches profile extends well beyond these predictions, with no evidence of King-like tidal radius out to $\sim$3 pc (Figure \ref{King_fit}).

To place a quantitative lower limit on the observed tidal radius, we use the Bayesian framework described above to fit our profile with a \citet{King:1962cr} model:

\begin{equation}
\label{eq:King}
\Sigma(r) = k * \left(\frac{1}{[1 + (r / r_c)^2]^{1/2}} - \frac{1}{[1 + (r_t / r_c)^2]^{1/2}}\right)^2 + b
\end{equation}
\\
where $k$ is a normalization constant, $b$ is a constant background term, and $r_c$ and $r_t$ are the core and tidal radii of the cluster, respectively. We adopt the core radius of 0.14 $\pm$ 0.05 pc measured by \citet{Espinoza:2009bs} as a prior for $r_c$, though our profile provides no additional information at R $\leq$ 0.25 pc. An uninformed prior is used for $r_t$, and $k$ is calculated such that the integral of the fitted profile yields the total number of cluster members observed after membership probability, completeness, and area corrections.

The result of the fit is summarized in Table \ref{KingModel} and marginalized posterior distributions presented in Appendix \ref{sec:posteriors_appendix}. We obtain a 3$\sigma$ lower limit of 2.8 pc on a King-like tidal radius for the cluster. Of course, it is quite possible that the Arches profile is truncated at all, and may behave as a power law throughout the full cluster extent. Regardless, it is clear that the cluster extends beyond its predicted tidal radius of 2.5 pc.

\begin{deluxetable*}{l l r}
\tablewidth{0pt}
\tabletypesize{\footnotesize}
\tablecaption{King Profile Fit Results}
\tablehead{
\colhead{Parameter\tablenotemark{a}} & \colhead{Prior\tablenotemark{b}} & \colhead{Result}
}
\startdata
$r_t$ (pc) & U(1, 15) & 2.8\tablenotemark{c} \\
$r_c$ (pc) & G(0.14, 0.05)\tablenotemark{d} & 0.13 $\pm$ 0.03  \\
$b$ (stars pc$^{-2}$) & U(0,15) & 3.46 $\pm$ 0.94  \\
$k$ (stars pc$^{-2}$) & --- & 1729 $\pm$ 643
\enddata
\label{KingModel}
\tablenotetext{a}{Description of parameters: $r_t$ = tidal radius; $r_c$ = core radius; $b$ = field contamination; $k$ Normalization factor}
\tablenotetext{b}{Uniform distributions: U(min, max), where min and max are bounds of the distribution; Gaussian distributions: G(mean, standard deviation)}
\tablenotetext{c}{3$\sigma$ lower limit}
\tablenotetext{d}{Source: \citet{Espinoza:2009bs}}
\end{deluxetable*}

\subsection{Tidal Breaks in the Radial Profile}
\label{Tidal_breaks}
We examine our profile for the presence of breaks which would be indicative of significant tidal interactions. The profile is remarkably consistent with a single power law from 0.25 pc $<$ R $<$ 1 pc, though beyond 1 pc the profile appears to exhibit slightly higher stellar densities than expected. This feature is adequately modeled as a constant field contamination term in Equation \ref{eq:PL_likelihood}. However, we must confirm that this feature can indeed be attributed to residual field contamination rather than a true over-density of cluster stars.

Field contamination may arise from the uncertainties in the cluster membership probabilities, which in turn are a result of uncertainties in the fits of the cluster and field kinematic distributions. To estimate the impact of these uncertainties, we perform a Monte-Carlo experiment where we randomly draw 1000 kinematic models from the joint posterior distribution obtained in $\mathsection$ \ref{sec:pops} and calculate the stellar cluster membership probabilities for each. We then determine the number of cluster members for each model from the sum of membership probabilities for all stars with P$_{member} >$ 0.3, the same calculation we do for the best-fit kinematic model in $\mathsection$ \ref{sec:Membership}. The standard deviation in the number of cluster members across the kinematic models is 34.32, centered around a median very nearly equal to the number of members identified by the best-fit kinematic model. We therefore adopt 34.32 as the uncertainty in the number of cluster members due to imperfect cluster membership probabilities.

The cluster membership uncertainty provides an estimate on the number of field contaminants potentially among our sample of cluster members. Spread evenly across the field, this would result in a field surface density of 1.49 stars pc$^{-2}$ in the profile. This is consistent with the background of the single power law fit (2.52 $\pm$ 1.32) to 1$\sigma$. In addition, Figure 1 of \citet{Penarrubia:2009pd} shows that a tidal event may cause an asymmetry in the parallel and perpendicular profiles, with the stellar over-density dominating the parallel profile relative to the perpendicular one. That no such asymmetry is observed ($\mathsection$\ref{sub:TT}) is further evidence that there is no tidal break in the Arches profile between 0.25 - 3.0 pc.

\section{Discussion}
In order to interpret the Arches cluster profile in terms of its tidal history, we require theoretical/numerical studies of YMCs on moderately eccentric orbits in the inner Milky Way potential that examine the evolution of the outer radial profile. If it exists, the observable King-like tidal radius of the cluster is significantly larger than predicted by previous studies, though these assume a spherically symmetric potential for the Galaxy or assume the cluster is on a circular orbit \citep{Kim:2000wd, Portegies-Zwart:2002hc}. Given the high gas densities toward the GC, the effect of interactions with Giant Molecular Clouds on the radial profile may also need to be considered. Unfortunately there are no such studies currently available in the literature; while there are many N-body simulations of clusters in varying tidal fields, very few discuss the corresponding evolution of the outer cluster profile. As a result, in $\mathsection$\ref{sec:Profile_implications} we consider the implications of simulation results by \citet[][hereafter P09]{Penarrubia:2009pd}, which model the response of dwarf spheroidal galaxies to tidal perturbations at perigalacticon. We discuss other simulations which more closely reflect the Arches cluster and its environment but do not examine the evolution of the radial profile in sufficient detail in $\mathsection$\ref{sec:otherStudies}.

\subsection{Time Since Last Pericenter Passage}
\label{sec:Profile_implications}
Modeling the effects of tidal stripping on dwarf spheroidal galaxies, P09 find a relation between the location of a tidal break in a radial profile and the time elapsed since perigalacticon. Here we apply this relation to the Arches cluster to place limits on the time elapsed since its closest approach to the GC. However, there is a major caveat in this analysis: the orbits examined by P09 are significantly different than what is expected for the Arches. These authors simulate dwarf spheroidal galaxies on highly eccentric orbits ($\epsilon$ = 0.96 -- 0.99) with a closest approach of 900 pc from the GC, where the Arches is likely on an orbit with $\epsilon$ = 0.25 -- 0.38 and a closest approach between 50 -- 200 pc from the GC \citep{Stolte:2008qy, Kruijssen:2015fx}. Assuming the Milky Way potential model described in \citet{Bovy:2015bd}, the radial force felt by the Arches is $\sim$6 times larger than that of the innermost P09 orbit. A similar formation and evolution of a tidal break in response to perigalacticon is found in N-body simulations of dwarf galaxies on more moderately-eccentric orbits (0.23 $<$ $\epsilon$ $<$ 0.9) by \citet{okas:2013hc}, though these models also probe weaker tidal fields than is experienced by the Arches (smallest pericenter: 12.5 kpc). N-body studies of star clusters on eccentric orbits show the formation of a tidal break at large radii, as well \citep{Kupper:2010ek, Johnston:1999qf, Lee:2006xr}. Given the supporting evidence from multiple studies and the lack of alternative studies in stronger tidal fields that examine the outer radial profile (see $\mathsection$\ref{sec:otherStudies}), we move forward with the results from the P09 simulations.

After perigalacticon, the radial profiles of P09 develop a tidal break that initially forms at small radii and moves outward over time. Normalizing by core radius R$_c$ and core crossing time t$_{cr}$~$\equiv$~R$_c$ / $\sigma_0$, they obtain the following relation:

\begin{equation}
\label{eq:breakRadius}
R_b / R_c = 0.55 (t - t_p) / t_{cr}
\end{equation}
\\
\noindent where R$_{b}$ is the radius of the tidal break and ($t - t_p$) is the time elapsed since perigalacticon. Given that no break is observed in the Arches profile between 0.25 pc -- 3.0 pc and adopting a velocity dispersion of 5.4 km s$^{-1}$ \citep{Clarkson:2012ty}, this relation indicates that the Arches has not had a significant tidal perturbation between $\sim$0.08 Myr and $\sim$1 Myr ago. However, P09 note that while their relation describes all models well at large break radii, it tends to underestimate the time since perigalacticon passage for break radii close to the core radius ( R$_b$ $\leq$ 4 R$_{c}$). Restricting the lower boundary to the innermost radius that is well described by this relation (0.6 pc for the Arches), we find that it is likely that the Arches has not experienced perigalacticon between $\sim$0.2 Myr and $\sim$1 Myr ago.

This result, which suggests that the cluster may either be nearly at or long past closest approach, places a limit on the set of potential orbits calculated for the Arches cluster \citep{Stolte:2008qy}. Restricting the orbits to those which place the Arches within the Central Molecular Zone (highly likely given its interactions with surrounding gas clouds, cf. \citeauthor{Wang:2010pr} 2010), this constraint rejects the viability of retrograde orbits since these place the cluster's closest approach within the timeframe in which we would expect to observe a tidal break. This provides further evidence that the cluster is on a prograde orbit and thus is located in front of the sky plane which passes through Sgr A*. However, we cannot significantly constrain the prograde orbits, which place the cluster very near closest approach. The allowed orbits are consistent with both cluster formation mechanisms discussed in $\mathsection$\ref{sec:intro}, and so no additional insight regarding the birth of the Arches cluster can be obtained.

Additional astrometric observations of the Arches are needed to provide higher precision proper motions from which the velocity dispersion profile of the cluster can be measured. Combined with the radial profile, the velocity dispersion profile is a sensitive tracer of cluster's tidal interaction history and current dynamical state, and may lead to the measurement of current Jacobi radius of the cluster \citep{Kupper:2010ek}. This can yield the present distance between the Arches and Sgr A*, the last bit of information required for a full orbital solution. The different formation mechanisms for the Arches can then be distinguished, as they differ in predictions of the current distance between the cluster and SMBH ($\sim$50 pc for the tidal compression scenario versus $\sim$100 - 200 pc for the cloud collision scenario).

It is important to note that the orbit calculations of the Arches cluster rely on the accurate measurement of the cluster's bulk motion relative to Sgr A*. This is not a trivial task due to an observational bias towards stars on the near side of the GC. Limited by their field of view, \citet{Stolte:2008qy} model the kinematics of 67 identified field stars using a single circular Gaussian distribution, taking the bulk motion of the Arches to be the difference between the kinematic centers of the Arches and field distributions (5.6 $\pm$ 0.5 mas yr$^{-1}$). This is the value used in the currently published orbits. Using 210 field stars and a more sophisticated elliptical Gaussian model for the field population, \citet{Clarkson:2012ty} obtain a slightly lower field motion of 4.39 $\pm$ 0.38 mas yr$^{-1}$ relative to the cluster. However, with a much larger field of view and more available field stars ($\sim$5322 stars with P$_{member} <$ 0.3), our study shows that the field has a complex kinematic structure that must be modeled with multiple Gaussians. Great care must be taken to properly interpret these structures in the context of a Galactic model and measure the bulk motion of the Arches with respect to Sgr A*, which is left to a future paper.

\subsection{Applicability of Other Theoretical Studies}
\label{sec:otherStudies}
Despite the significant caveats in applying the P09 results to the Arches cluster, it is the most applicable theoretical study currently available that examines the detailed evolution of the outer radial profile of a stellar system as it passes through perigalacticon. N-body simulations of dwarf galaxies on orbits with moderate eccentricities by \citet{okas:2013hc} exhibit tidal breaks that behave similarly as those in the P09 models, though the relation between the location of the break radius and the time since perigalacticon passage is not assessed quantitively. Alternatively, N-body simulations by \citet{Kupper:2010ek} examine objects more similar to the Arches, studying the behavior of 10$^4$ M$_{\odot}$ star clusters on elliptical orbits with 0.25 $<$ $\epsilon$ $<$ 0.70 (smallest pericenter: 600 pc). The resulting cluster profiles show that extratidal stars form power-law extensions at large cluster radii, though the detailed evolution of the profile after perigalacticon is not explored in detail. Additional simulations of clusters on elliptical orbits by \citet{Johnston:1999qf} and \citet{Lee:2006xr} also show extratidal stars forming a break in the cluster profile, though neither study examines how this break evolves as a function of orbital phase.

There is a large body of additional literature studying the dynamical evolution of star clusters in tidal fields, though these do not show the evolution of the outer radial profile. N-body simulations of clusters on eccentric orbits by \citet{Baumgardt:2003qy}, \citet{Lamers:2010kn}, and \citet{Webb:2014ul} primarily focus on the evolution of the mass-loss rate and mass function, while \citet{Webb:2013ee} examines the half-mass radius, tidal radius, and cluster size rather than the morphology of the radial profile at large radii. Many other studies examine clusters on circular orbits, though these are limited to old ($>$10 Gyr) globular clusters \citep{Trenti:2010dz, Gieles:2011ys}, or do not present detailed radial profiles of their models \citep{Ernst:2009pb, Madrid:2012zt}.

It is worthwhile to note that none of the studies discussed above examine the evolution clusters within the central regions of the Milky Way. There have been several studies of Arches-like young compact clusters within the inner 200 pc of the Galaxy, but these similarly do not examine the evolution of the outer radial profile. N-body simulations by \citet{Kim:2000wd} and \citet{Portegies-Zwart:2002hc} make predictions regarding expected cluster lifetimes and the evolution of the radial profile out to the half-mass radius. \citet{Portegies-Zwart:2004rc} focus on the evolution of the mass function of Arches-like clusters, concluding that the mass function of the inner region of the Arches reported by \citet{Figer:2002nr} could be explained by dynamical mass segregation. Several other studies model the effects of dynamical friction on compact clusters, predicting in-spiral towards the GC and their subsequent evolution, though the cluster profile during this process is not presented \citep{Kim:2003bl, Portegies-Zwart:2003oe, Gurkan:2005sw}. Additional studies of the behavior of the radial profiles of Arches-like clusters near the GC are required to draw more conclusive interpretations from the observations presented here.

\section{Conclusions}

We have conducted a multi-epoch photometric and astrometric study of the Arches cluster using the \emph{Hubble Space Telescope} WFC3IR camera at 1.27, 1.39, and 1.53 $\mu$m. Using a sophisticated astrometric pipeline we extract individual stellar proper motions to an accuracy of at least 0.65 mas yr$^{-1}$ down to F153M $\approx$ 20 mag ($\sim$2.5$_{\odot}$), reaching a precision of $\sim$0.1 mas yr$^{-1}$ for the brightest stars. Taking advantage of the distinct kinematic properties of the cluster, we use a 4-Gaussian mixture model to simultaneously fit the cluster and field proper motion distributions and calculate cluster membership probabilities. This is a substantial improvement over photometrically-determined cluster membership due to the large degree of differential reddening across the field. The field of view in this study is 144 times larger than previous astrometric studies of the Arches cluster, allowing for the identification of high-probability cluster members out to a cluster radius of 75" ($\sim$3 pc at 8 kpc).

Combining the cluster membership probabilities, an extinction map derived from red clump (RC) stars, and an extensive completeness analysis, we construct the stellar radial density profile for the Arches cluster between 6.25" $<$ R $<$ 75" (0.25 pc $<$ R $<$ 3.0 pc) down to a differentially de-reddened magnitude of F153M = 20 mag. This profile is well fit by a single power-law of slope $\Gamma$ = 2.06 $\pm$ 0.17 with a constant field contamination density of 2.52 $\pm$ 1.32 stars pc$^{-1}$. Surprisingly, no evidence of a tidal radius is observed. Adopting a King profile as a model, we obtain a 3$\sigma$ lower limit of 2.8 pc for the observed tidal radius of the Arches cluster. This shows that the cluster extends beyond its largest predicted theoretical tidal radius of 2.5 pc.

Additionally, we examine the Arches cluster profile for evidence of mass segregation and tidal tails. We find the cluster to exhibit mass segregation at all observed radii, with the radial profile of bright stars (F153M $<$ 17 mag, or M$>$ $\sim$13 M$_{\odot}$) being notably steeper than the profile of fainter (17 $<$ F153M $<$ 20 mag) stars. A KS test reveals the differences between these profiles to be significant. We leave a careful conversion from brightness to mass for a future paper. Similarly, we search for evidence of tidal tails by comparing the profile parallel to the direction of orbit to the profile perpendicular to it. No statistically significant asymmetries are observed in these profiles, as would be expected from tidal tail structures. Further observations, perhaps at larger cluster radii, are needed to continue to search for tidal tails.

No evidence of a tidal break is observed in the radial profile, as might be expected if the Arches has experienced a tidal perturbation in its recent past. Assuming that the results of dynamical simulations of dwarf spheroidal galaxies on highly eccentric orbits by \citet{Penarrubia:2009pd} can be applied to the Arches, this suggests that the Arches not likely experienced its closest approach to the GC within 0.2 -- 1 Myr ago. This constraint would reject all possible retrograde orbits of the cluster, providing further evidence that the Arches is on a prograde orbit and located in front of the sky plane which intersects Sgr A*. However, additional simulations studying the profile of Arches-like clusters on mildly-eccentric orbits in the inner Milky Way potential are required to interpret the observed profile with higher confidence.

Further astrometric observations of the Arches to obtain its velocity dispersion profile are needed to better constrain its orbit and distinguish between different possible cluster formation scenarios. It is important to note that an accurate determination of the cluster's orbit requires measuring its bulk motion with respect to Sgr A*, which is a difficult task given the complex kinematic structure of the field population revealed in this study. A revised measurement of the Arches proper motion, along with a new calculation of possible orbits, is left to a future paper.

\acknowledgments
The authors thank the anonymous referee for insightful comments which improved this paper and acknowledge support from HST GO-13809. A.M.G. is supported by the NSF grants AST-0909218 and AST-1412615, and the Lauren Leichtman \& Arthur Levine Chair in Astrophysics. This work is based on observations made with the NASA/ESA Hubble Space Telescope, obtained at the Space Telescope Science Institute, which is operated by the Association of Universities for Research in Astronomy, Inc., under NASA contract NAS 5-26555. These observations are associated with programs \#11671, 12318, and 12667. This research has made extensive use of NASA's Astrophysical Data System. MWH would also like to acknowledge the SWOOP writing retreat and its participants for useful feedback.

\bibliography{ms_bib}

\clearpage

\clearpage
\appendix

\section{WFC3IR Measurements and Proper Motions}
\label{sec:Obs_methods_appendix}

In this appendix we describe the methods and software used to extract high precision astrometry, photometry, and proper motions from the WFC3IR observations. This is the first application of this methodology on WFC3IR observations. Stars are first detected and measured in the \texttt{flt} images using the FORTRAN program \emph{img2xym\_wfc3ir\_stdpsf} developed by Jay Anderson. Similar to the code \emph{img2xym\_WFC.09x10} documented in \citet{Anderson:2006il}, this performs PSF-fitting measurements using a library of spatially-variable PSF models it derives for the WFC3IR camera. This library contains a 3x3 grid of PSFs that spans the camera's field, where the PSF at any point can be derived from a spatial interpolation of these models. Since this routine operates on one image at a time in a single pass, it is not designed to deal with overlapping stars. It reduces each star as if it is the only contribution to the 5x5 pixels centered on its brightest pixel.

Each image is run through this program twice. The first iteration extracts the bright high S/N stars in the field ($\sim$400 stars per image), simultaneously measuring the residuals between the observed PSF and the library PSF. New image-specific PSF models are created in order to minimize these residuals and make them uniform across the field. The second iteration then uses the modified PSF library to accurately measure both bright and faint sources, producing a star list with fluxes and positions for $\sim$13,000 sources in each image extending down to F127M = 22.45 mag, F139M = 22.09 mag, and F153M = 21.71 mag. A small number of stars ($\sim$200) with F153M $\leq \sim$15 are saturated and are thus measured using the outer part of the PSF.

Next, we cross-identify stars from each exposure with a master list for the filter/epoch set, initially taken to be the first image in the set. Common stars found in at least 75\% of the images are used to transform positions from the distortion-corrected frame of each exposure into the master list reference frame using general 6-parameter linear transformations. This gives $\sim$N observations for each star, where N is the number of images in the filter/epoch set, which allows us to find a robust average position and flux. These averaged measurements produce an overall star catalog for the filter/epoch. We then adopt these catalogs as the new reference-frame positions and repeat the procedure, improving the transformations. The improvement in the second iteration is considerable, decreasing astrometric residuals by nearly a factor of 2. Finally, the new star catalogs for each filter/epoch are then transformed to an arbitrary astrometric reference frame where the net motion of the cluster plus field stars is 0 mas yr$^{-1}$. This produces what we will call the ``one-pass" catalogs, because they are limited to the stars which can be detected in a single image.

\begin{figure*}
\begin{center}
\includegraphics[scale=0.35]{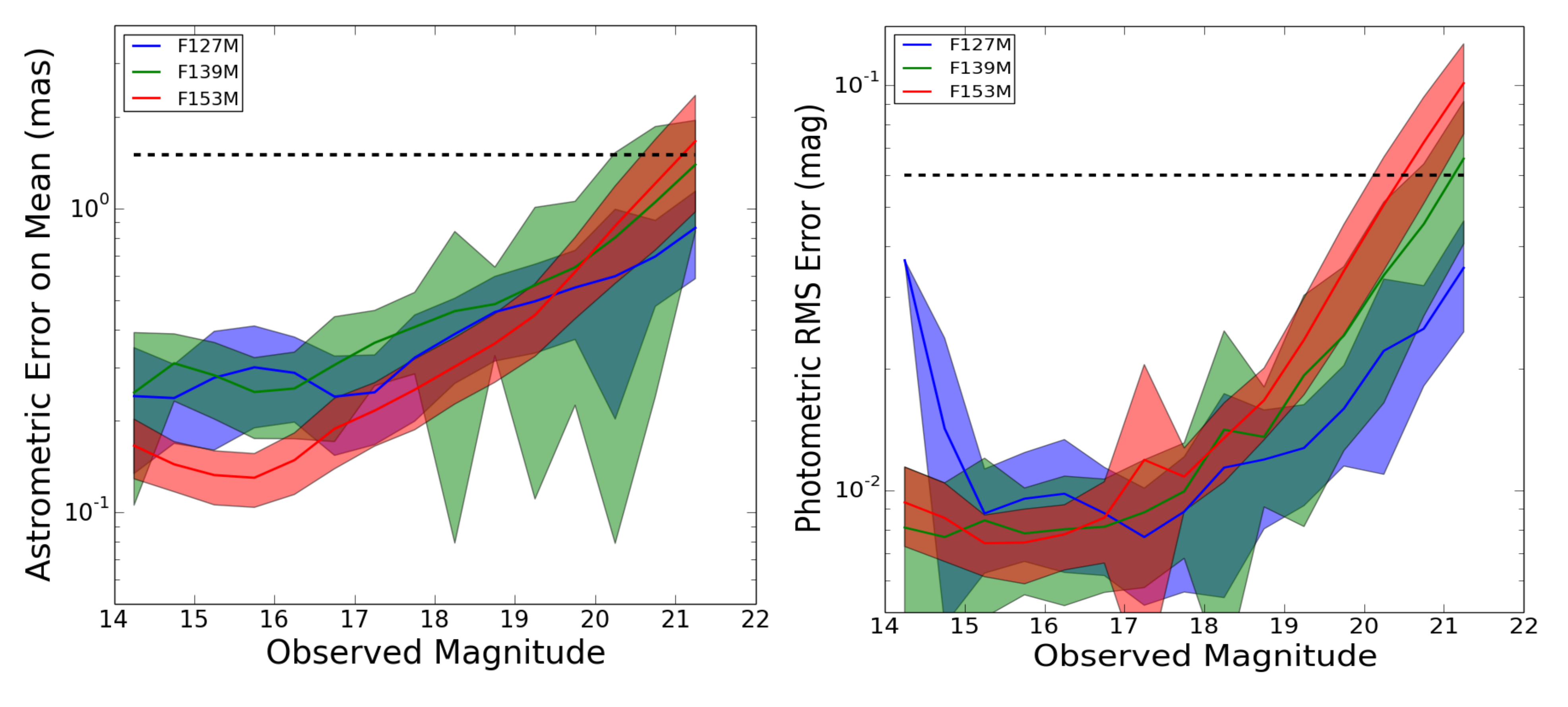}
\caption{Astrometric error on the mean and photometric RMS error vs. observed magnitude for the F127M (blue), F139M (green), and F153M (red) filters. The solid lines show the median errors and the shaded regions cover one standard deviation. Stars with astrometric errors above 1.5 mas (and thus proper motion errors above 0.65 mas yr$^{-1}$) or photometric errors above 0.06 mag in the F153M filter are not included in the analysis. These cuts are shown by the black dotted lines.}
\label{err_allfilters}
\end{center}
\end{figure*}

In principle, we could continue our analysis of the Arches Cluster using the one-pass catalogs. However, by stacking the images in each filter/epoch we significantly increase the detection depth, important because signal from otherwise undetected faint stars can be mistakenly associated with brighter stars and introduce biases in the measurements. With the image transformations, PSF models, and one-pass catalogs as input, we use the program \emph{KS2} to stack the images, make stellar detections to significantly fainter magnitudes, and then redo the astrometric and photometric measurements in each individual image at the positions found in the stacked image. The measurements for the individual images are then averaged together to produce the final catalogs for each epoch/filter. We emphasize that we don't use the measurements of the stacked image, but only those of the individual images. This iterative multiple-finding strategy increases the number of stars detected to $\sim$50,000 per filter (nearly five times as many as the one-pass analysis), reaching F127M~= 23.63 mag, F139M = 23.29 mag, and F153M =  23.31 mag as noted in $\mathsection$ \ref{sec:obs}. To ensure the accuracy of our transformation we again combine the individual star lists into star catalogs for each filter/epoch and re-align to the common astrometric reference frame using a first-order bi-variate polynomial (see \citeauthor{Ghez:2005tx} 2005 and references therein). Higher order fits were found to introduce artificial structure in the astrometry. Astrometric and photometric errors as a function of magnitude are presented in Figure \ref{err_allfilters}.

\begin{figure}
\begin{center}
\includegraphics[scale=0.4]{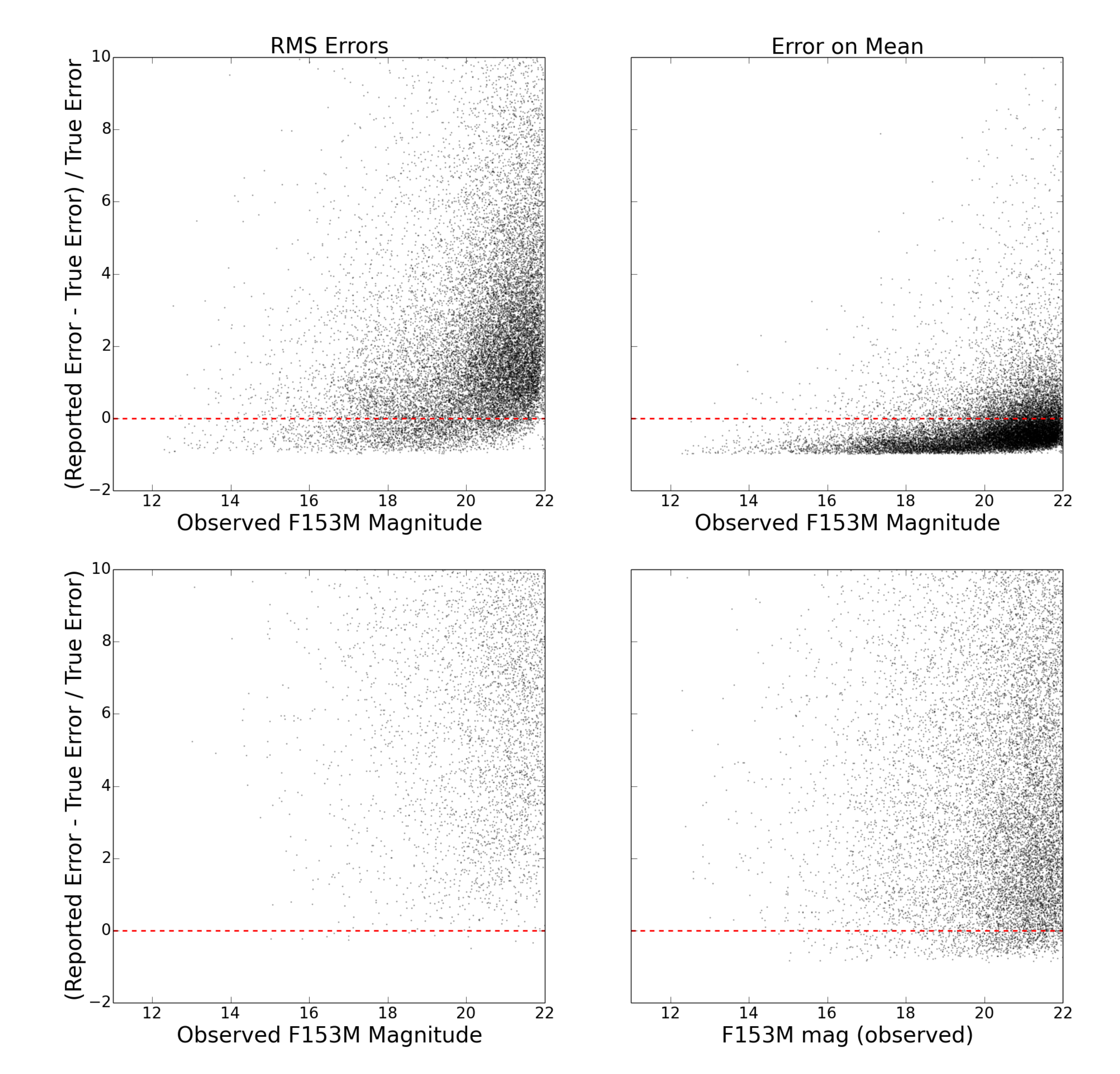}
\caption{\emph{Top:} The fractional difference between the RMS error (\emph{left}) and error on the mean (\emph{right}) and the ``true'' errors for the photometric measurements. The ``true" photometric error for each star is taken to be the standard deviation of the observed magnitude across the F153M epochs. \emph{Bottom:} Similar to above, the fractional difference between the RMS error (\emph{left}) and error on the mean (\emph{right}) and the ``true'' errors for the astrometric measurements. The ``true'' errors are taken to be the RMS residuals between the fitted proper motion and the observed position of each star in the F153M epochs. For the photometry, the mean is found to underestimate the true error, and so the photometric RMS error is adopted for individual measurements. For the astrometry, the RMS error is found to strongly overestimate the true error, and so the astrometric error on the mean is adopted for individual measurements.}
\label{RMSvSTD}
\end{center}
\end{figure}

The \emph{KS2} code returns a final star catalog with the root-mean-square errors ($\sigma_{RMS}$) for the astrometric and photometric measurements for each star in the filter/epoch. Theoretically the errors should be quantified by the error on the mean ($\sigma_{RMS}$ / $\sqrt{N_{obs}}$, where $N_{obs}$ is the number of images in the filter/epoch). To test this, we compare the \emph{KS2} errors to quantities measured across the F153M epochs that are directly caused by these errors: the standard deviation of the magnitude of each star (assuming no stellar variability) and the residuals between the measured position and position predicted by the star's proper motion. This comparison reveals the error on the mean to better capture the astrometric errors and the RMS error to better capture the photometric errors, and so we adopt these for the individual star measurements throughout (Figure \ref{RMSvSTD}). This choice does not significantly affect the proper motion errors described below, which are dominated by the position residuals rather than the astrometric errors themselves.

Proper motions are derived independently for the X and Y directions (in the image coordinate system) using a linear fit to the change in position over the F153M epochs:

\begin{equation}
x = x_{0} + v_x (t - t_{0})
\end{equation}
\begin{equation*}
y = y_{0} + v_y (t - t_{0})
\end{equation*}
\\
where $t_{0}$ is the astrometric error-weighted average time of observations, ($v_{x}, v_{y}$) are the X and Y proper motions, and ($x_0$, $y_0$) and ($x$, $y$) are the star positions at $t_0$ and $t$, respectively. To test the validity of these errors, we apply the derived proper motions to their respective stars and examine the residuals between the predicted and the observed positions. The distributions of the X and Y position residuals are approximately Gaussian, though more power is present in the wings than expected (Figure \ref{resid}). This is a consequence of stellar crowding distorting our measurements, as these wings are dominated by faint stars (F153M $<$ 20 mag) which are more prone to this effect. Given the high stellar density of near the center of the cluster this is not unexpected. The distribution of $\chi^2$ values for the proper motion fits follows the expected distribution for 1 degree of freedom, appropriate as we constrain 2 parameters with 3 measurements.

\begin{figure*}
\begin{center}
\includegraphics[scale=0.35]{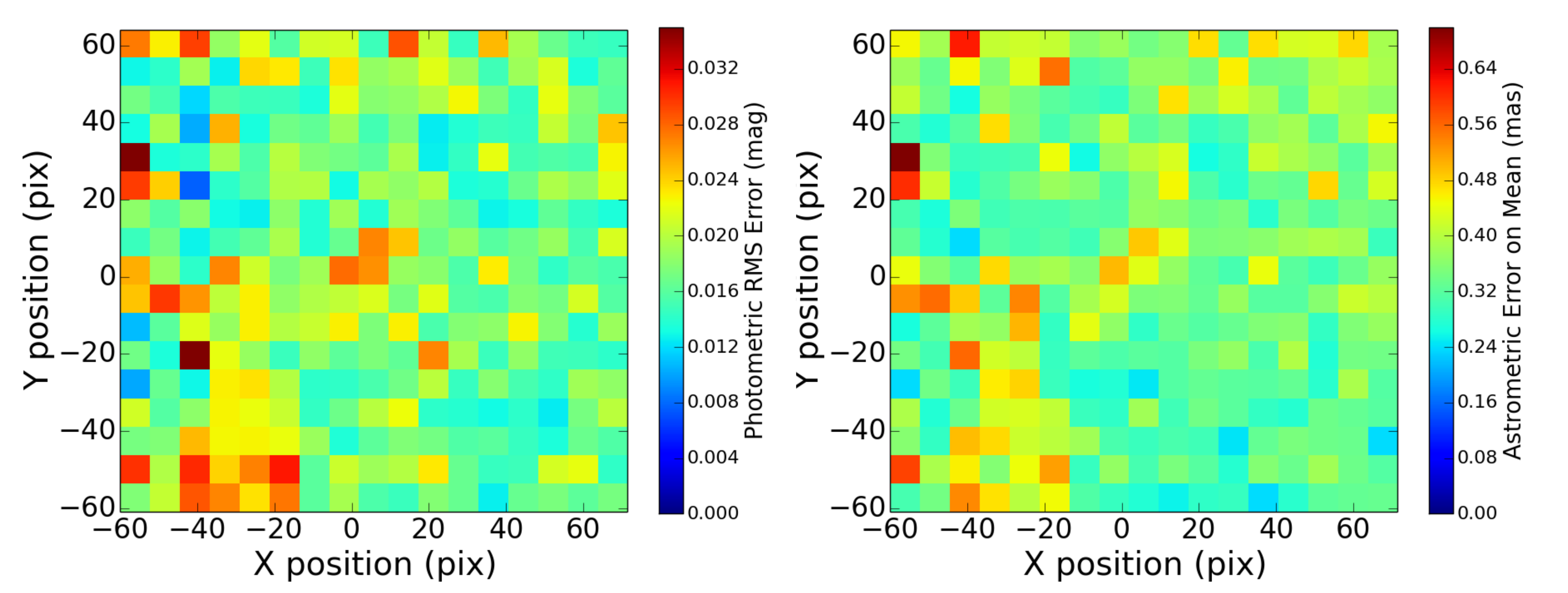}
\caption{Average astrometric error (\emph{left}) and F153M photometric error (\emph{right}) as a function of position on the camera for all stars with F153M $<$ 20 mag. Average error values are calculated in 7'' bins and are plotted relative to the cluster center. Axes are oriented in the same manner as Figure \ref{Arches_color}. Higher errors are observed in the low reddening regions and the cluster center due to stellar crowding.}
\label{spatial_errs}
\end{center}
\end{figure*}

\begin{figure}
\begin{center}
\includegraphics[scale=0.4]{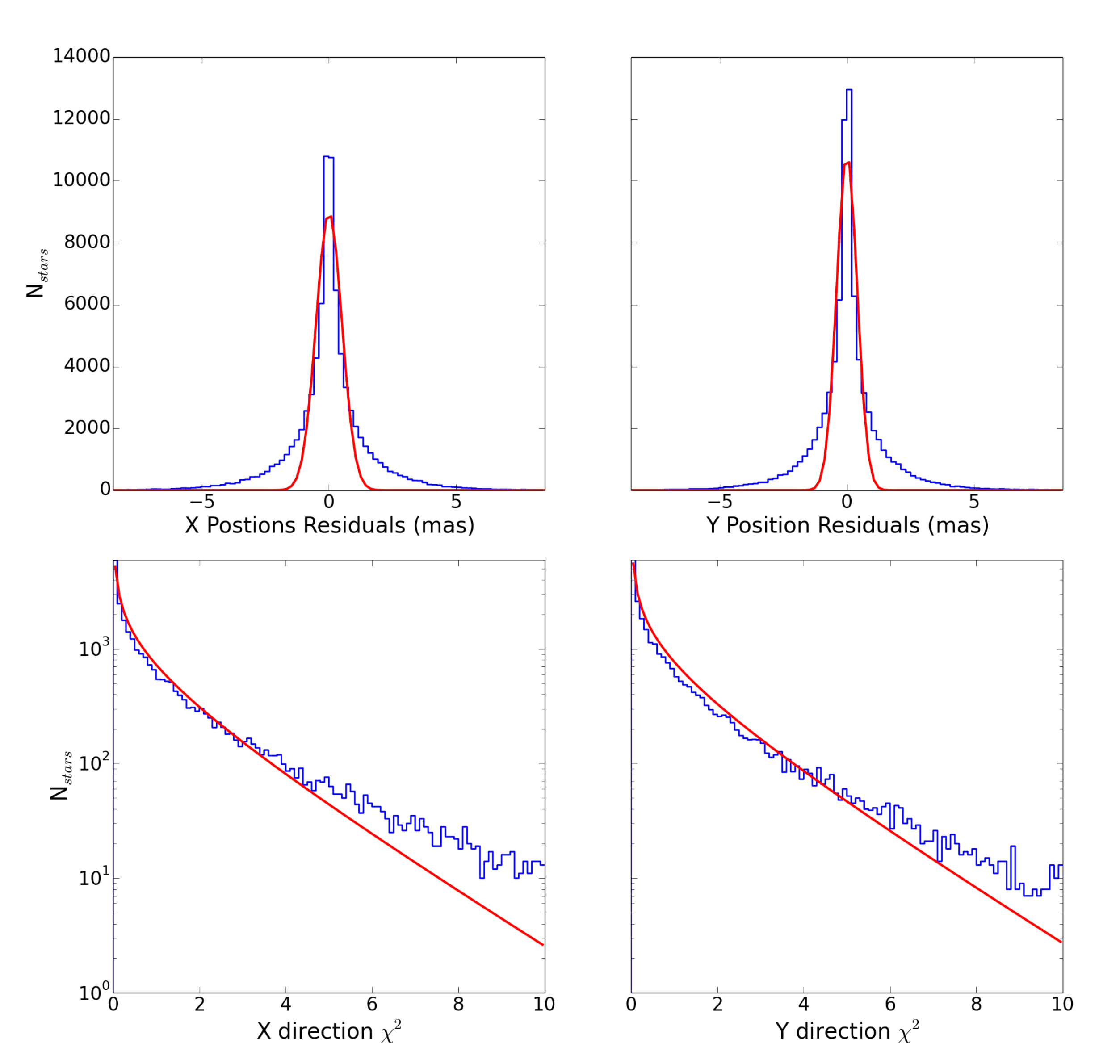}
\caption{Positions residuals between the observed positions and those predicted from the fitted proper motions. \emph{Top:} A histogram of the X (\emph{left}) and Y (\emph{right}) residuals (blue line). Ideally these distributions would be Gaussian (red line), but stellar crowding results in more power in the wings of the distribution. \emph{Bottom:} The distribution of $\chi^2$ values for the proper motion fits in both X (\emph{left}) and Y (\emph{right}), where we adopt the error on the mean as the proper motion error (blue line). These values match the expected $\chi^2$ distribution with 1 degree of freedom (red line), validating our reported errors.}
\label{resid}
\end{center}
\end{figure}

Photometry is calibrated to the standard Vega magnitude system by deriving the filter-dependent offset between \emph{KS2} magnitudes and 0.4'' aperture photometry magnitudes. Aperture photometry is performed using \emph{DAOPHOT} \citep{S87} on the \texttt{drz} image for each epoch/filter, a composite image of all exposures in the same filter/epoch produced by the \emph{HST} pipeline. This offset is then combined with the appropriate 0.4" zeropoint derived for the WFC3IR camera (see $\mathsection$ \ref{sec:obs} for reference) to determine the overall zeropoint for the \emph{KS2} observations. For stars with F153M $\leq$ 20 (consistent with our proper motion precision cut, $\mathsection$ \ref{sec:obs}), the median F153M astrometric and photometric errors are 0.34 mas and 0.018 mags, respectively, with evidence of higher errors in regions of increased stellar density (Figure \ref{spatial_errs}). Within this sample the median F127M and F139 photometric errors are 0.033 mag and 0.025 mag, respectively.

\section{Artificial Star and Observed Star Errors}
\label{sec:Artstar_appendix}

\begin{figure}
\begin{center}
\includegraphics[scale=0.4]{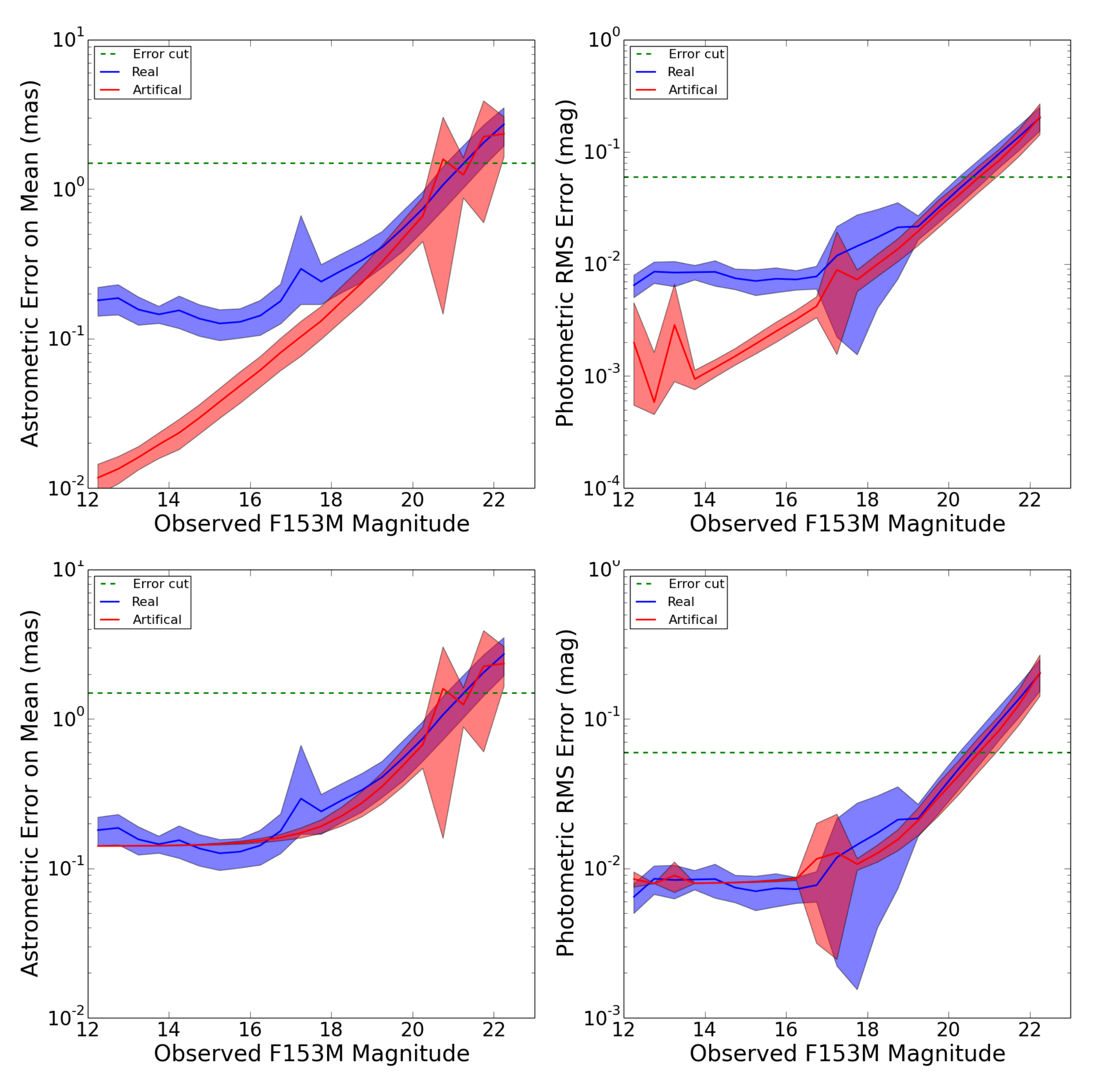}
\caption{Comparison of the astrometric (\emph{left}) and photometric (\emph{right}) errors for the observed and artificial stars in blue and red, respectively. The \emph{top} plots show the initial artificial star errors and the \emph{bottom} plots show the artificial star errors after a constant error term has been added. The solid line and filled area represents the median and standard deviation of the errors. The additional error terms reflect the error floor seen in the observed star measurements, and are measured to be of 0.14 mas and 0.008 mag for the astrometry and photometry, respectively. After this adjustment the error distributions match well. The green dotted line shows the error cuts used in this study.}
\label{errcomp_artobs}
\end{center}
\end{figure}

The completeness analysis described in $\mathsection$ \ref{sec:complete} assumes that the measured artificial star errors match the observed star errors. A direct comparison of the errors reveals that the observed astrometric and photometric errors have an error floor that is not reproduced by the artificial star tests (Fig. \ref{errcomp_artobs}). A possible explanation for this feature is residual PSF variations which are not captured by our spatially-varying PSF model, as the artificial stars are planted using this model and thus wouldn't reflect this error. A constant error correction term of 0.14 mas and 0.008 mag is added to the artificial star position and magnitude uncertainties, after which the artificial star errors are found to closely follow those of the observed stars. This error correction does not have a noticeable effect on the completeness analysis as only stars far below the error cuts are significantly affected.

\section{Posteriors of Profile Fits}
\label{sec:posteriors_appendix}

\begin{figure*}
\begin{center}
\includegraphics[scale=0.3]{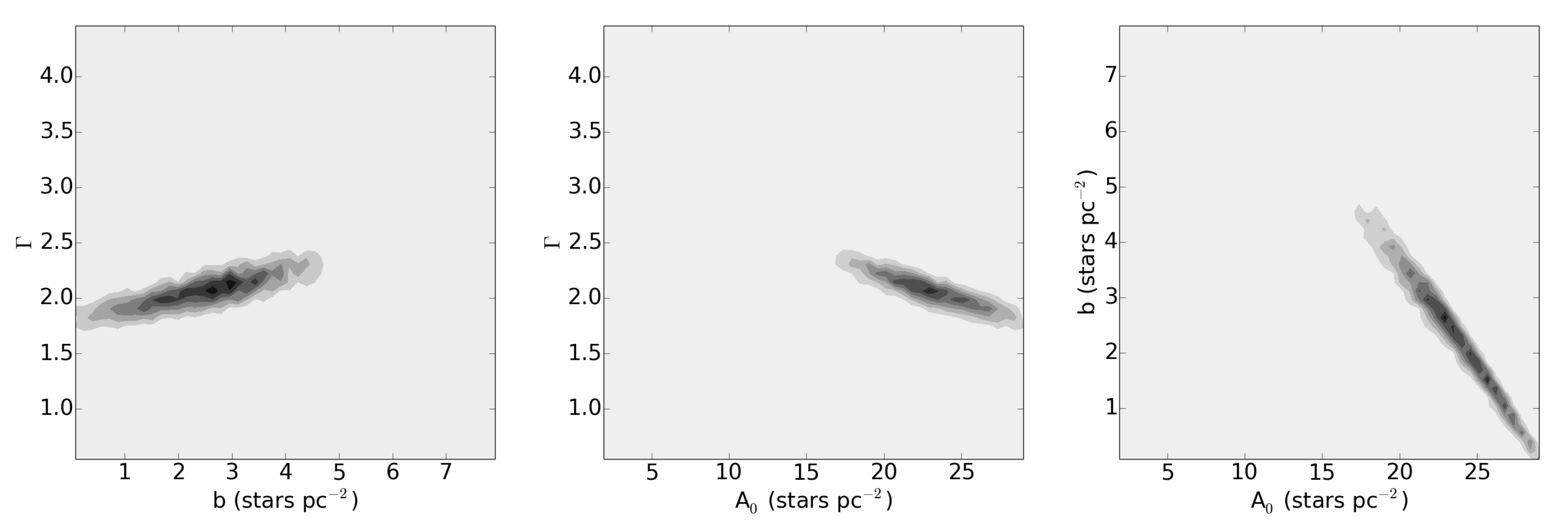}
\caption{The bivariate posterior distributions for the power law model fit to the full cluster sample. }
\label{PL_posteriors}
\end{center}
\end{figure*}

\begin{figure*}
\begin{center}
\includegraphics[scale=0.3]{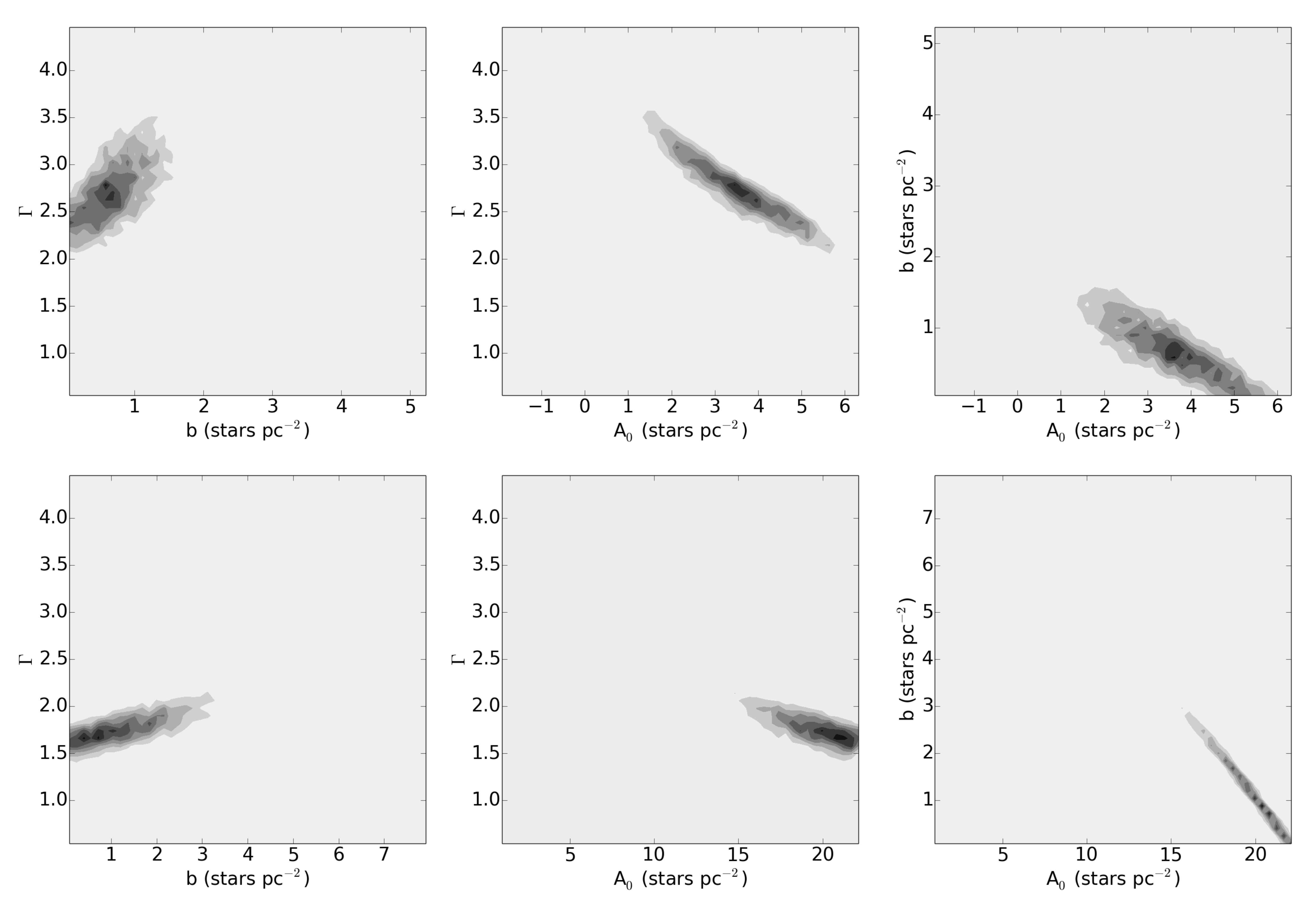}
\caption{The bivariate posterior distributions for the power law model fits to the high-mass (\emph{top}, F153M $<$ 17 mag) and low-mass (\emph{bottom}, F153M $>$ 17 mag) cluster members. That the high-mass stars have a steeper power law slope $\Gamma$ is an indication of mass segregation in the Arches cluster. }
\label{MS_posteriors}
\end{center}
\end{figure*}

\begin{figure*}
\begin{center}
\includegraphics[scale=0.25]{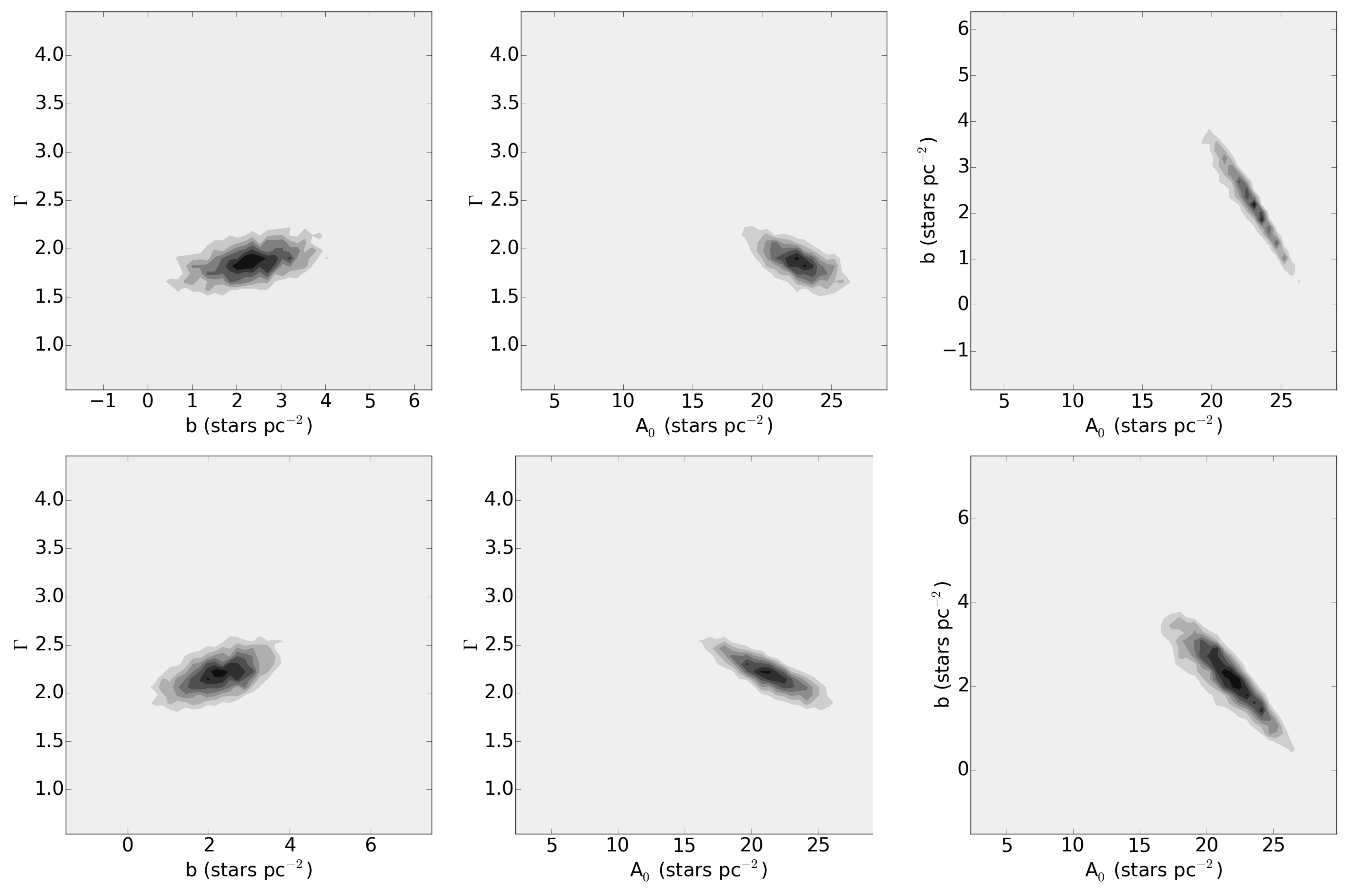}
\caption{The bivariate posterior distributions for the power law model fits to the cluster members parallel (\emph{top}) and perpendicular (\emph{bottom}) to the bulk cluster orbit. The existence of tidal tails would cause asymmetries in these profiles such as a difference in the power law slope $\Gamma$ of these profiles. No significant evidence for tidal tails are found.}
\label{TT_posteriors}
\end{center}
\end{figure*}

In this appendix we present the bivariate posterior distributions for the different power law profile fits described in $\mathsection$\ref{sec:profile} and the marginalized and bivariate posterior distributions for the King profile fit described in $\mathsection$\ref{sec:tidal}. The bivariate posterior distributions show the correlations between the power law slope ($\Gamma$), background level ($b$) and profile amplitude ($A_0$) for the profile fits to the full cluster (Figure \ref{PL_posteriors}), high-mass/low-mass members (Figure \ref{MS_posteriors}), and the parallel/perpendicular members (Figure \ref{TT_posteriors}). The marginalized posteriors of these parameters (not shown) are well described by Gaussians, which provide the best fit parameter values and errors reported in Table \ref{PL_results}.

\begin{figure*}
\begin{center}
\includegraphics[scale=0.5]{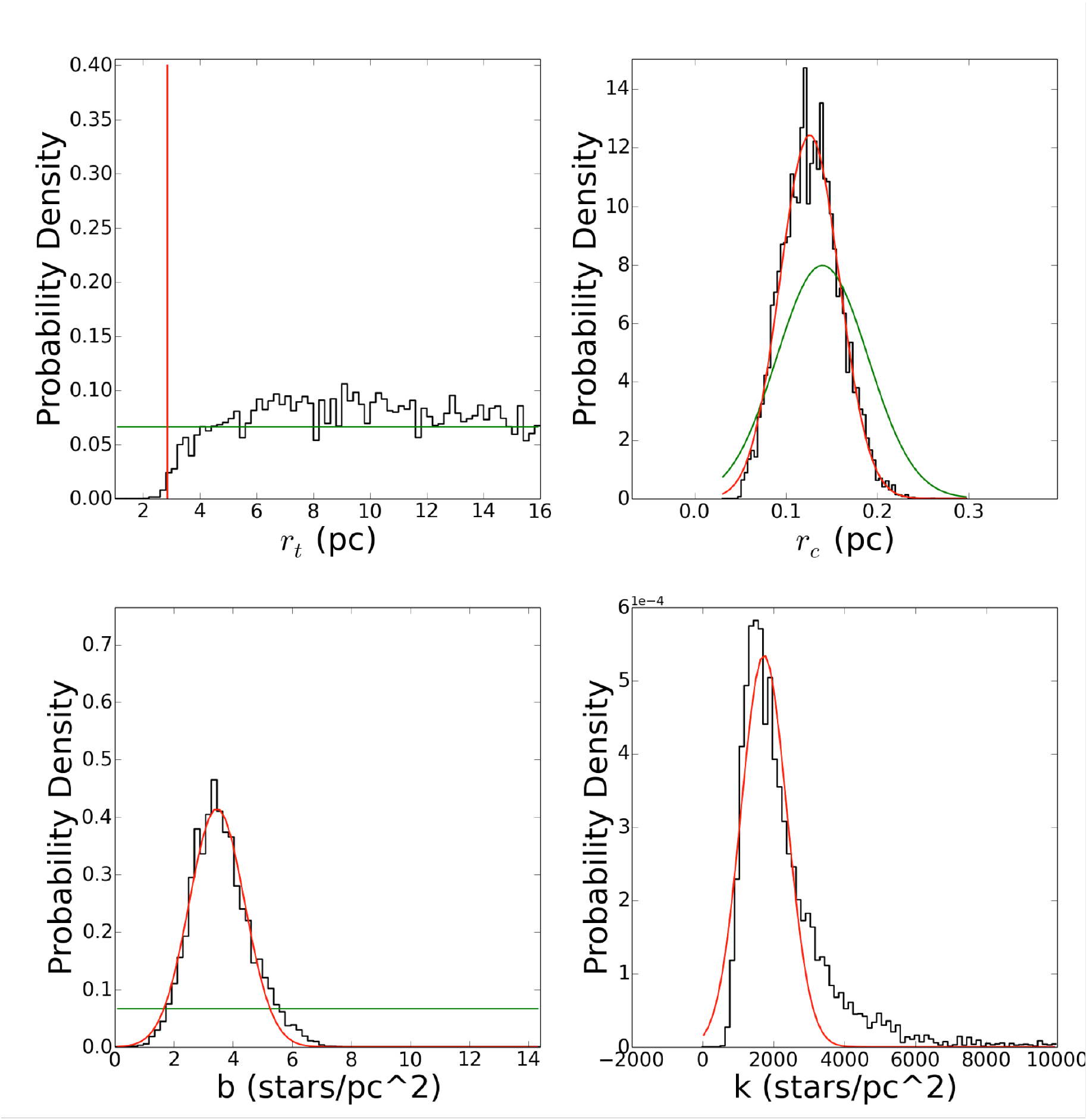}
\caption{The 1D marginalized posterior distributions for the King model fit. The output of the \emph{Multinest} sampling is in black, the corresponding Gaussian fit in red, and the input prior in green (if applicable). The 3$\sigma$ limit to the tidal radius $r_t$ (2.8 pc) is indicated by the red line, where 99.7\% of the best-fit models fall above this value.}
\label{King_posteriors}
\end{center}
\end{figure*}

\begin{figure*}
\begin{center}
\includegraphics[scale=0.25]{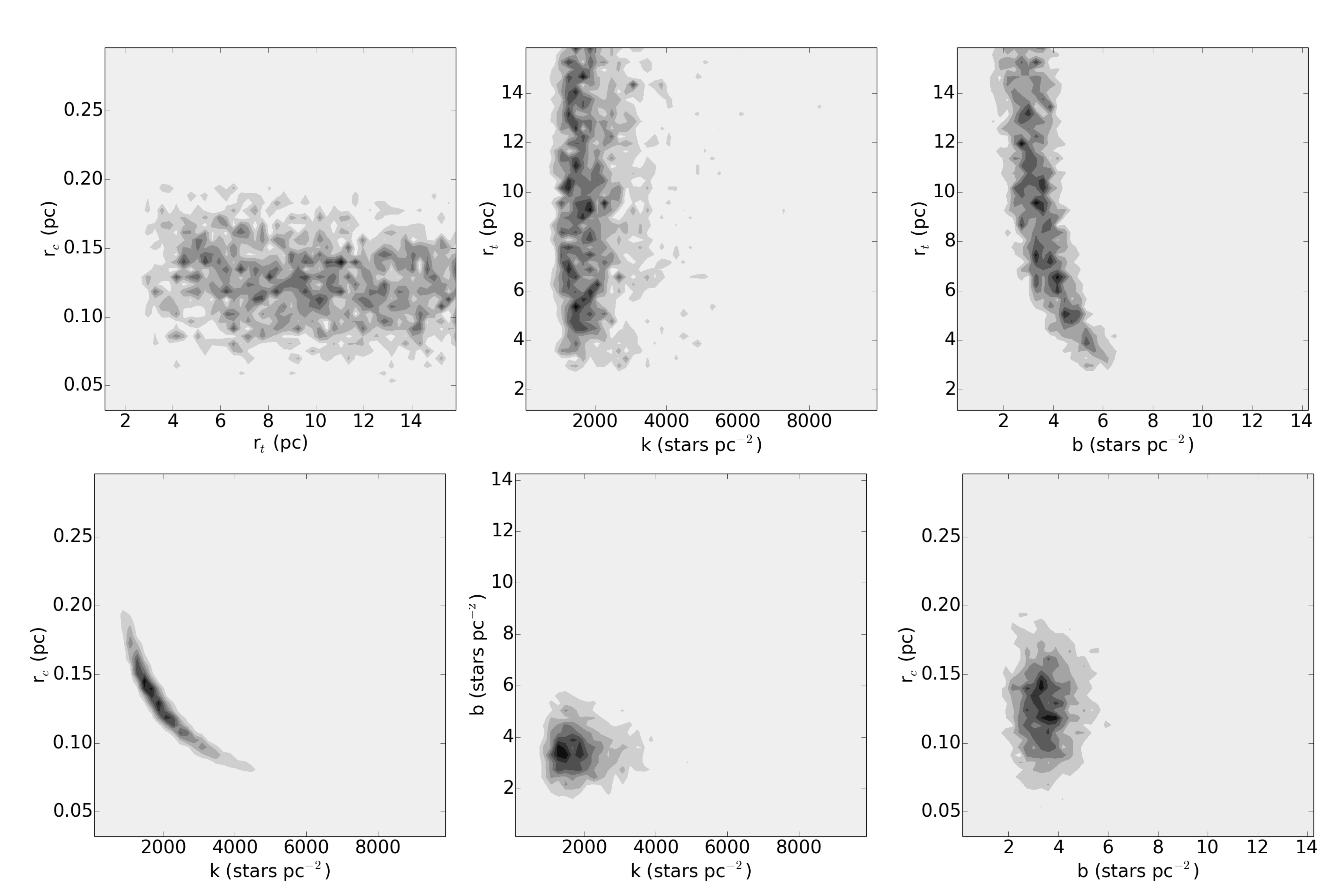}
\caption{The bivariate posterior distributions for the King model fit.}
\label{King_posteriors2D}
\end{center}
\end{figure*}

For the King model fit, the marginalized posterior distribution for the tidal radius (r$_{t}$), core radius (r$_c$), background ($b$), and normalization factor ($k$) are shown in Figure \ref{King_posteriors}, while the bivariate posterior distributions are shown in Figure \ref{King_posteriors2D}. The significance of the 3$\sigma$ lower limit on the tidal radius is discussed in $\mathsection$\ref{sec:tidal}. The best fit parameter values and errors are reported in Table \ref{KingModel}.

\end{document}